\documentclass[aps,preprint,nofootinbib]{revtex4}

\usepackage{graphicx}
\usepackage{multirow}
\usepackage{amsmath,amssymb}
\usepackage{url}

\usepackage[normalem]{ulem}
\usepackage{xcolor}

\newcommand{\lsim}{\mathrel{\mathop{\kern 0pt \rlap
  {\raise.2ex\hbox{$<$}}}
  \lower.9ex\hbox{\kern-.190em $\sim$}}}
\newcommand{\gsim}{\mathrel{\mathop{\kern 0pt \rlap
  {\raise.2ex\hbox{$>$}}}
  \lower.9ex\hbox{\kern-.190em $\sim$}}}
\newcommand{\mev}{{\,{\rm MeV}}}
\newcommand{\gev}{{\,{\rm GeV}}}
\newcommand{\tev}{{\,{\rm TeV}}}
\newcommand{\al}{{\alpha}}

\newcommand{\Gm}{{\Gamma}}

\newcommand{\Th}{{\Theta}}

\newcommand{\beq}{\begin{equation}}
\newcommand{\eeq}{\end{equation}}
\newcommand{\bea}     {\begin{eqnarray}}
\newcommand{\eea}     {\end{eqnarray}}
\newcommand{\no}{{\nonumber}}

\newcommand{\on}{ {(1)} }
\newcommand{\tw}{ {(2)} }

\newcommand{\zt}{ { Z^{(2)} }}
\newcommand{\lo}{ { L^{(1)} }}

\newcommand{\lt}{ { L^{(2)} }}

\newcommand{\lkp}{ {B^{(1)}}}

\newcommand{\neu}{\tilde{\chi}^0}

\newcommand{\neuo}{{\tilde{\chi}^0_1}}

\newcommand{\mmaxcaso} { m^{\rm max}_{\rm cas1}}
\newcommand{\mcuspcaso} { m^{\rm cusp}_{\rm cas1}}
\newcommand{\mtmaxcaso} { (m_T)^{\rm max}_{\rm cas1}}
\newcommand{\mtcuspcaso} { (m_T)^{\rm cusp}_{\rm cas1}}

\newcommand{\ptmaxcaso} { (p_{T})^{\rm max}_{\rm cas1}}
\newcommand{\mmaxcast} { m^{\rm max}_{\rm cas2}}

\newcommand{\mtcuspcast} { (m_T)^{\rm cusp}_{\rm cas2}}

\newcommand{\ptncuspcast} { (p_{Tn})^{\rm cusp}_{\rm cas2}}
\newcommand{\ptnmaxcast} { (p_{Tn})^{\rm max}_{\rm cas2}}

\newcommand{\tyo} {\textsc{Type~I}}
\newcommand{\tyt} {\textsc{Type~II}}
\newcommand{{\mao}}{{{\sc Mass--a}$_1$}}
\newcommand{{\mat}}{{{\sc Mass--a}$_2$}}
\newcommand{{\mbo}}{{{\sc Mass--b}$_1$}}
\newcommand{{\mbt}}{{{\sc Mass--b}$_2$}}
\newcommand{{\mco}}{{{\sc Mass--c}$_1$}}
\newcommand{{\mct}}{{{\sc Mass--c}$_2$}}

\begin{document}

\preprint{PITT-PACC 1204}

\title{Kinematic Cusps 
with Two Missing Particles II: \\
Cascade Decay Topology}
\author{Tao Han$^{1}$, Ian-Woo Kim$^2$,  Jeonghyeon Song$^{3}$}
\affiliation{
$^1$ Pittsburgh Particle physics, Astrophysics, and Cosmology Center, Department of Physics $\&$ Astronomy, University of Pittsburgh, 3941 O'Hara St., Pittsburgh, PA 15260, USA\\
$^2$Department of Physics, University of Michigan, USA\\
$^3$Division of Quantum Phases \& Devises, School of Physics, 
Konkuk University,
Seoul 143-701, Korea }
%
%
\begin{abstract}
Three-step cascade decays into two invisible particles and two visible particles via two intermediate on-shell particles develop cusped peak structures in 
several kinematic distributions.
We study the basic properties of the cusps and endpoints
in various distributions and demonstrate 
that the masses of the missing particles and the intermediate particles can be determined by the cusp and endpoint positions. Effects from realistic considerations such as finite decay widths, longitudinal boost  of the parent particle, and spin correlations are shown to be under control for the processes illustrated. 
\end{abstract}
\maketitle

\section{Introduction}
\label{sec:introduction}

At the energy frontier, the LHC experiments are taking us to an unprecedented territory 
of the Tera-scale physics beyond the Standard Model (SM). 
At the cosmo frontier, we have entered an era of precision cosmology. 
With much progress made in the two frontiers, 
we have to admit that
our understanding of the Universe is still far from being complete. 
According to the precise measurements of the 
cosmic microwave background fluctuations,  
such as WMAP~\cite{wmap},
about 95\% component of the current universe
has never been directly detected in the laboratory.
The dominant component ($\approx 72\%$) is 
dark energy that is responsible for the
accelerating expansion 
of the universe~\cite{dark:energy}.
The second dominant ($\approx 23\%$) is 
cold dark matter (CDM), which is assumed to be in a form of nonrelativistic matter but cannot be explained within
the SM.
Albeit its extraordinary success in explaining current experimental data
with incredibly high precision,
the SM is regarded as an effective
theory below a certain scale. 
For example, theoretical unnaturalness of 
the SM, dubbed as {\em gauge hierarchy problem}, 
suggests new physics beyond the SM at the TeV scale.
Therefore, it is a very intriguing possibility that such CDM components 
may appear in new physics models.

Indeed, some new particle physics models
have answers for the astrophysical question about CDM. 
One of the most popular scenarios is a thermal production of 
weakly interacting massive particles (WIMP)~\cite{WIMP}. In this scenario,  
 a stable particle $X$ had been  once 
in thermal equilibrium in the early history of the universe,
but got frozen out as its reaction rate became slower than the
expansion of the universe.
The stability of the CDM particle over cosmic time
is often due to an unbroken parity symmetry
or a discrete symmetry.
Under such a symmetry, 
the SM particle fields are 
in the trivial representation while
new particle fields are in some nontrivial representation.
The decay of the lightest new particle
into SM particles is prohibited.
The current observation  
highly suggests that 
the CDM particle has its mass at the electroweak scale 
and its couplings 
with a size of weak interaction.
Some popular models with WIMPs are 
supersymmetric models with $R$ parity~\cite{SUSY:DM},
the universal extra dimension (UED) model  
with Kaluza-Klein (KK) parity~\cite{DMExtraD},
and
the littlest Higgs model with $T$ 
parity~\cite{DMLittleHiggs}.

This WIMP with an electroweak scale mass
is likely to
 be produced at the LHC. 
In hadron colliders, such weakly 
interacting neutral particles can be identified only by missing 
transverse energy.   
The measurement of its mass is of crucial importance
to reveal the identity of the CDM, but 
this is a very challenging task at the LHC 
because such invisible particles are pair-produced with
combinatoric complications and with large errors especially in jet energy 
measurements. 
In the literature, many new ideas 
to measure the CDM mass have been proposed~\cite{Burns:Review},
such as endpoint methods~\cite{endpoint},
polynomial methods~\cite{polynomial:combine,polynomial},
and $M_{T2}$ 
methods~\cite{MT2:original,MT2,MT2:kink,subsystem:MT2}.

\begin{figure}
\begin{center}
\includegraphics[scale=1]{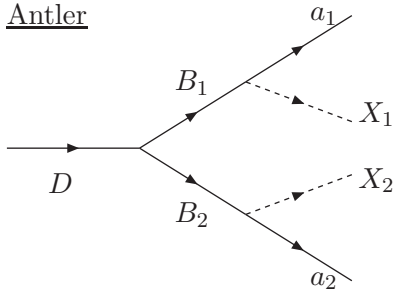}
\end{center}
\caption{
The antler decay topology of a parity-even particle $D$
into two missing particles ($X_1$ and $X_2$) and two visible 
particles ($a_1$ and $a_2$).
}
\label{fig:feyn:ant}
\end{figure}

Recently, we have proposed a new approach to measure the missing particle mass
by using the singular structures in the kinematic
distributions of the antler decay~\cite{cusp,long:antler}.
The antler decay is a resonant decay of 
a parity-even particle $D$ into a pair of the missing particles
($X_1$ and $X_2$)
and a pair of SM visible particles ($a_1$ and $a_2$)
through two on-shell parity-odd intermediate particles ($B_1$ and $B_2$),
as depicted in Fig.~\ref{fig:feyn:ant}.
We have studied two kinds of singular structures, 
an endpoint and a cusp.
The positions of cusps and endpoints
determine the masses of the
missing particle as well as the intermediate particle,
if the parent particle mass $m_D$ is known 
from other decay channels directly 
into two SM particles\footnote{This is possible 
since the particle $D$ has \emph{even} parity.}.

There are  
a few interesting merits of this method:
(i) the positions of the cusp and endpoint
are stable under
the spin correlation effects since it is purely 
determined by the phase space;
(ii) a cusp as a sharp and non-smooth 
\emph{peak} is 
statistically more advantageous to search than
endpoints, and more identifiable to observe than kinks;
(iii) the simple configuration
of outgoing particles 
can reduce combinatoric complication
which is commonly troublesome
in many missing particle mass measurement methods;
(iv)
the derived analytic functions for some
kinematic distributions 
are very useful to reconstruct
the mass parameters by best-fitting.

\begin{figure}
\centering
  \includegraphics[scale=1]{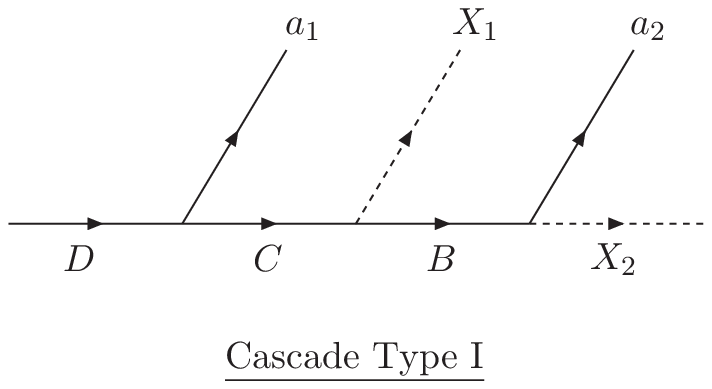}
  \phantom{x}
  \includegraphics[scale=1]{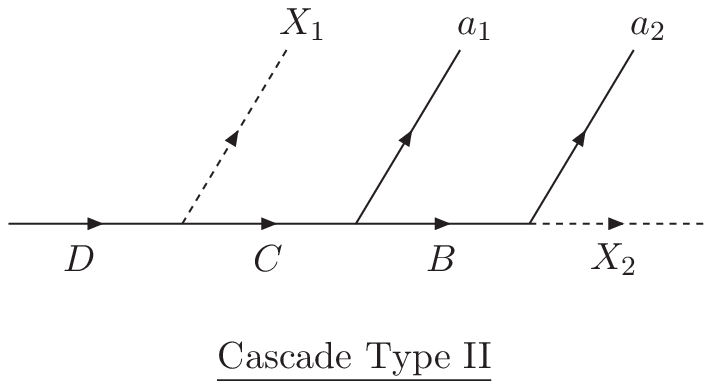}
  \caption{
The cascade decay topology of a parity-even particle $D$
into two missing particles ($X_1$ and $X_2$) and two visible 
particles ($a_1$ and $a_2$).}
\label{fig:feyn:cas}
\end{figure}

In this paper, as a companion of Ref.~\cite{long:antler}, 
we focus on another decay topology with 
two visible particles and two missing particles from a parity-even 
particle $D$: {\em cascade} decays shown in Fig.~\ref{fig:feyn:cas}.
In this process, the parent 
particle $D$  
sequentially decays into two particles through three steps in series,
finally ended up with a missing particle $X_2$. 
There are two non-trivial types of three-step 
cascade decay,
according to at which step the 
first missing particle $X_1$ is produced.
\tyo~and \tyt~cascade decays have different cusp and endpoint structures.
Unlike the symmetric
antler decay case with one kind of intermediate particle,
the cascade decay involves two different intermediate 
particles.
We thus
need to fix one more unknown mass,
which requires more independent observables.
The study of the basic properties of cusp and endpoint
in various kinematic distributions
to determine the unknown masses for the three-step
cascade decay is our main purpose.
The cusp in the invariant mass distribution of \tyo~decay
has been discussed in
the context of new physics models with 
the CDM particle stabilized by $Z_3$ symmetry~\cite{Z3}.

The rest of the paper is organized as follows.  
In Sec.~\ref{sec:kin},
we categorize all possible kinematic variables from 
the four-momenta of the two visible particles.
Section \ref{sec:type1} deals with the \tyo~cascade decay.
We present the expressions of cusps and endpoints
of various kinematic distributions in a common case
where $m_{a_1} = m_{a_2} =0$
and $m_{X_1} = m_{X_2}$.
The functional form of the invariant mass distribution
is also given.
The general mass case is to be discussed in the Appendix.
In Sec.~\ref{sec:type2},
we present the corresponding results for the \tyt~cascade decay.
Section \ref{sec:effects} is devoted to realistic considerations
such as the finite widths of the intermediate particles,
the longitudinal boost of the parent particle $D$, and the
spin correlation.
We then conclude in Sec.~\ref{sec:conclusions}. 

\section{Kinematics of cascade decay topology with 
two missing particles}
\label{sec:kin}

We consider the four-body cascade decay of a heavy particle $D$
through three consecutive steps.
The cascade decay resulting in
a \emph{single} missing particle and three visible particles
has been extensively studied in the literature.
In the MSSM, a good example is the process of $\tilde{q} \to q \neu_2
\to q \ell_n \tilde{\ell} \to q \ell_n\ell_f \neuo$.
In the UED model, we have
$Q^\on \to Z^\on q \to L^\on \ell_n q \to B^\on \ell_f \ell_n q$.
Here $\ell_n (\ell_f)$ denotes the near (far) lepton 
with respect to the parent particle.
In principle, three observable particles
provide enough information to determine
all the unknown mass parameters 
involved~\cite{Burns:Review,endpoint}.
However, there are some difficulties in extracting
proper information, especially because of combinatoric complications.
It is hard to 
distinguish $\ell_n$ from $\ell_f$.
Furthermore, the parent 
particle $D$ is to be pair-produced due to
its odd parity (or nontrivial representation), and thus 
there is always another decay chain in the same event.

Here we consider the three-step cascade decay with two missing 
particles.
The parent particle $D$ is 
of \emph{even} parity and thus
its single production is allowed.
The final states are simply two visible particles ($a_1$
and $a_2$) with missing transverse
energy.
There is no combinatoric complication
when forming the invariant mass of two visible particles.
In addition, if the rest frame of $D$ in the transverse direction
can be determined, 
the individual transverse momenta of 
$a_1$ and $a_2$ in the frame also show kinematic singularities which 
have additional information on the mass parameters of the system.  
Note that this decent feature relies on the information of 
$D$'s transverse motion. 

The cascade decays of $D\to a_1 a_2 X_1 X_2$
can be classified 
according to at which step the 
first missing particle, say $X_1$, is produced.
Note that we fix the other missing particle ($X_2$) 
to be produced at the last step.
If $X_1$ is also from the last step,
the final intermediate particle $B$ is just missing
and this decay is 
indistinguishable from a two step decay. 
We do not consider this case.
Then, there are two non-trivial three-step 
cascade decays,
as depicted in Fig.~\ref{fig:feyn:cas}.
In the \tyo~decay, $X_1$ is from the second step.
The parent particle $D$ decays into a visible particle $a_1$
and a new particle $C$,
followed by the decay of $C$ into a missing particle $X_1$
and a new particle $B$.
Finally $B$ decays into a visible particle $a_2$ and
a missing particle $X_2$.
In the \tyt~decay, $X_1$ is from the first step:
$D$ decays into $C X_1$,
followed by $C\to a_1 B$, and  finally $B\to a_2 X_2$.
In the view point of two observable particles $a_1$ and $a_2$, 
this \tyt~decay
is a two-step cascade decay 
of a new heavy particle $C$.
As shall be shown, there is no cusp structure in 
Lorentz-invariant distributions. 

It is useful to describe the kinematics of the three-step cascade decay 
in terms of the rapidity of individual massive particles or a combination
of multiple particles: 
\begin{equation}
\eta^{(k)}_{i} = {E^{(k)}_{i} \over m^{(k)}_{i} } ,
\end{equation}
where $E_{i}$ and $m_{i}$ are the energy and mass of the particle (system) $i$
in the rest frame of a particle (system) $k$. 
To avoid confusion, we adopt the following rapidity notations 
for the \tyo~and \tyt~decays: 
\bea
\no
\begin{array}{c|c|c}
 & \hbox{ \tyo~Cascade} &
\hbox{ \tyt~Cascade} \\ \hline
\hbox{ rapidity notation } & \xi_i & \zeta_i
\end{array}
\eea
For the sake of simplicity, when the rapidity 
is defined in the rest frame of its mother particle,
we omit the superscript specifying the reference frame. 

With the four-momenta $k_1$ and $k_2$
of the two observable particles $a_1$ and $a_2$ in the lab frame,
respectively, we consider the following 
observables in three categories:
\begin{itemize}
\item Lorentz invariant observables: the invariant mass of $a_1$ and $a_2$, 
 \bea
\label{eq:m:def}
m = \sqrt{(k_1+k_2)^2} \ .
\eea
\item Longitudinal-boost invariant observables:
\begin{itemize}
\item the magnitude of the transverse momentum of a visible particle $i$, 
\bea
\label{eq:pTi:def}
p_{Ti} =\left| \mathbf{k}_i^{T} \right| ,
\eea
\item the magnitude of the transverse momentum of the $a_1$-$a_2$ system, 
\bea
\label{eq:pT:def}
p_{T} = \left| \mathbf{k}_1^{T} + \mathbf{k}_2^{T}\right| ,
\eea
\item the transverse mass of the $a_1$-$a_2$ system, 
\bea
\label{eq:mT:def}
m_{T} = \sqrt{
p_{T}^{2 }+ m^2} \ .
\eea
\end{itemize}
\item Non-invariant observables:
\begin{itemize}
\item cosine of
$\Theta_i$, the angle of the visible particle $a_i$
in the c.m.~frame of $a_1$ and $a_2$, with respect to their c.m.~moving direction,
\bea
\label{eq:cos:def}
\cos\Theta
 = \frac{\mathbf{k}_1^{(aa)} \cdot \mathbf{k}^{(D)}}
 { |\mathbf{k}_1^{(aa)} | |\mathbf{k}^{(D)}| } \ .
\eea
\end{itemize}
\end{itemize}
Here the bold-faced letter denotes the three-vector momentum,
$k = k_1 + k_2$, and the superscript $(D)$
and $(aa)$ denote the $D$-rest frame and the c.m. frame of $a_1$ and $a_2$,
respectively.

As shall be shown, $p_{Ti}$ and $m_{T}$ distributions show
cusp structures if the mother particle $D$ is produced at rest 
in the transverse direction.
At a hadron collider, this is possible if $D$ is singly produced.
These additional cusp structure are very valuable to determine all the unknown masses.
However strong QCD interactions always yield, for example, sizable initial
state radiation,
which causes transverse kick to the mother particle $D$:
the cusps in $p_{Ti}$ and $m_{T}$ distributions can be smeared.
Caution is required when drawing the consequences for the mass measurement
from these cusps.
In addition, the $\cos\Theta$ distribution is defined in the rest-frame of the mother particle $D$.
At a hadron collider, this is not observable.
In what follows, we assume that the mother particle is
produced at rest in the transverse direction.

In general, the involved seven particles 
($D$, $C$, $B$, $a_{1}$, $a_2$, $X_{1}$, $X_2$)
may have different masses.  
In many new physics models, 
the cascade decay processes 
of interest
have massless visible particles and the same kind of invisible particles. 
For most of the presentation in the main text,  therefore,
we consider only the following case:
\bea
\label{eq:realistic:cascade}
m_{a_1} = m_{a_2} = 0,
\quad
m_{X_1} = m_{X_2}.
\eea
The result for the most general case with seven different masses is 
presented in the Appendix.

\section{\tyo~cascade decay}
\label{sec:type1}

As illustrated in Fig.~\ref{fig:feyn:cas},
the \tyo~cascade decay is
the decay of a parity-even particle $D$
into two missing particles $X_1$ and $X_2$
and two visible particles $a_1$ and $a_2$ through
\bea
D(P) \longrightarrow  && C + a_1 (k_1),       \\
\no
          && C \longrightarrow       B + X_1 ,  \\ \no
          &&  \hspace{34pt}
 B \longrightarrow      a_2(k_2) + X_2.
\eea
Here the particles $D$, $C$, $a_1$ and $a_2$
are parity-even while the particles $B$, $X_1$, and $X_2$
are parity-odd.
In order to accommodate the \tyo~cascade decay,
we need at least two heavy parity-even particles.

One good example for
this decay channel 
is in the universal extra dimension (UED) model~\cite{ued}.
It is based on a single flat extra dimension of size $R$,
compactified on an $S_1/Z_2$ orbifold.
All the SM fields propagate freely in the whole five-dimensional spacetime,
and each field has an infinite number of KK 
excited states.
Since the KK parity is conserved,
the lightest KK particle (LKP) with odd KK parity is stable
and becomes a good candidate for the CDM. 
Usually the first KK mode of the $U(1)_Y$ gauge boson 
$\lkp$ is the LKP~\cite{ued,radiative:ued}. 
All the second KK states 
of the SM particles have even KK-parity and
mass of $\sim 2/R$.
Lower limit of $1/R \gsim 300$ GeV is set based on the combination
of the constraints from the $\rho$ parameter \cite{rho},
the electroweak precision tests \cite{EWPT},
the muon $g-2$ measurement \cite{gmuon},
the flavor changing neutral currents \cite{fcnc},
and direct search by D0 group at the Tevatron \cite{d0}.
The second KK modes are within the reach of the LHC.
Possible \tyo~cascade decays are
\bea
\label{eq:UED:Z2}
\zt &\to& \ell_n +\lt  \to \ell_n  + \lkp\lo
\to \ell_n + \lkp +\ell_f\lkp,
\\ \label{eq:UED:g2}
g^\tw & \to& q_n + q^\tw \to q_n + \lkp q^\on
\to q_n+ \lkp + q_f \lkp.
\eea

Now we present the cusps and endpoints
of $m$, $m_T$, $p_T$, $p_{Ti}$, and $\cos\Th$
distributions in terms of the masses.
As in Eq.~(\ref{eq:realistic:cascade}), 
two missing 
particles are of the same kind 
and the visible particles are massless
in this case.   
The rapidities of the particles $B$ and $C$ 
in the rest frame of their mother particles are
given by
\bea
\label{eq:cas1:rapidity}
\cosh \xi_B
= \frac{m_C}{2 m_B} \left( 1+\frac{m_B^2}{m_C^2} -
\frac{m_X^2}{m_C^2} \right), \quad \cosh \xi_C = \frac{m_D}{2 m_C}
\left( 1+\frac{m_C^2}{m_D^2} \right) .
\eea
We will also use $E_n$ and $E_f$, the energy of the near
$a_1$ and the far $a_2$ in its mother's rest frame, respectively:
\bea
\label{eq:En:Ef}
E_n = \frac{m_D}{2}\left( 1-\frac{m_C^2}{m_D^2} \right),
\quad
E_f = \frac{m_B}{2}\left( 1-\frac{m_X^2}{m_B^2} \right).
\eea

\begin{table}[t!]
\centering
\begin{tabular}{|c||c|c|c|c||c||}
\hline
         & ~~$m_D$~~ & ~~$m_C$~~ & ~~$m_B$~~&~~$m_X$~~
            & ~~~$\xi_B$~~~ ~~ \\
\hline
~~~{\sc Mass--a}$_1$~~~  & 1045.7  & 1023. &514.2 & 500.9 
& 0.12  \\
~~~{\sc Mass--b}$_1$~~~  & 600 & 400 & 200 & 100 & 
0.60 \\
~~~{\sc Mass--c}$_1$~~~  & 600 & 500 & 150 & 100 & 
1.16 \\
\hline
\end{tabular}
\caption{ Test mass spectrum sets for the \tyo~cascade decay. 
All masses are in units of GeV. }
\label{table:cascade1:mass}
\end{table}

For illustration,
we take three sets for the mass parameters 
in Table \ref{table:cascade1:mass}.
The {\sc Mass--a}$_1$
is motivated by the $Z^{(2)}$ decay in Eq.~(\ref{eq:UED:Z2}).
The KK masses are determined by
the UED model parameters of $\Lambda R=20$ and $1/R=500\gev$,
where $\Lambda$ is
the cutoff scale~\cite{ued}.
The equal spacing of the KK mode spectrum in flat extra 
dimension  
leads to very degenerate masses, \textit{i.e.},
$m_D \approx m_C \approx 2 m_B \approx 2 m_X$.
The {\sc Mass--b}$_1$ has substantial  gaps
for each pair of adjacent masses.
Finally {\sc Mass--c}$_1$ has a sizable mass gap between
$m_C$ and $m_B + m_X$.

For precise mass measurements using the singularities, 
it is required to have a visible cusp and/or endpoint in a kinematic 
distribution. 
The visibility of the endpoints 
can be determined by
the functional behavior near the endpoint.
In what follows, 
the shape of an endpoint is to be classified into
a fast dropping one and a long-tailed one.

\vskip 0.5mm
\noindent
\underline{\textit{(i) Invariant mass $m$ distribution}:}
We first discuss the distribution of the invariant mass $m$
of two visible particles.
The differential decay rate  
$d \Gm/dm$ 
is
\begin{eqnarray}
\frac{d\Gamma}{dm} 
\propto
\begin{cases}
\begin{array}{ll}
 2 \xi_B m, &  
\hbox{ for } 0 < m < \mcuspcaso, \\
m \,\ln \dfrac{(\mmaxcaso)^2}{m^2} ,
& 
\hbox{ for } \mcuspcaso < m < \mmaxcaso ,
\end{array}
\end{cases}
\label{eq:cas1:cusp:eq}
\end{eqnarray}
where the cusp and endpoint are
\bea
\label{eq:cas1:cusp:max}
\left(
\mcuspcaso
\right)^2
=
4 E_n E_f e^{\xi_C-\xi_B},
\quad \left(
\mmaxcaso
\right)^2 = 4 E_n E_f e^{\xi_C+\xi_B}.
\eea
Note that the functional behavior of $d \Gm/d m$
is the same as that of the antler decay~\cite{long:antler}.
More general case with 7 different masses is discussed in the Appendix.

Whether this cusp is sharp enough to probe
can be easily deduced from Eq. (\ref{eq:cas1:cusp:eq}).
The $d \Gm/ d m $ function is linear in $m$ for $m< \mcuspcaso$,
and a concave function for $\mcuspcaso < m <\mmaxcaso$.
At $m=\mmaxcaso/e$,
the concave function reaches its maximum.
If $\mmaxcaso/e < \mcuspcaso$,
which is equivalent to
$\xi_B<1$,
the cusp can be considered to be pronounced.

\begin{figure}[!t]
\centering
  \includegraphics[scale=0.8]{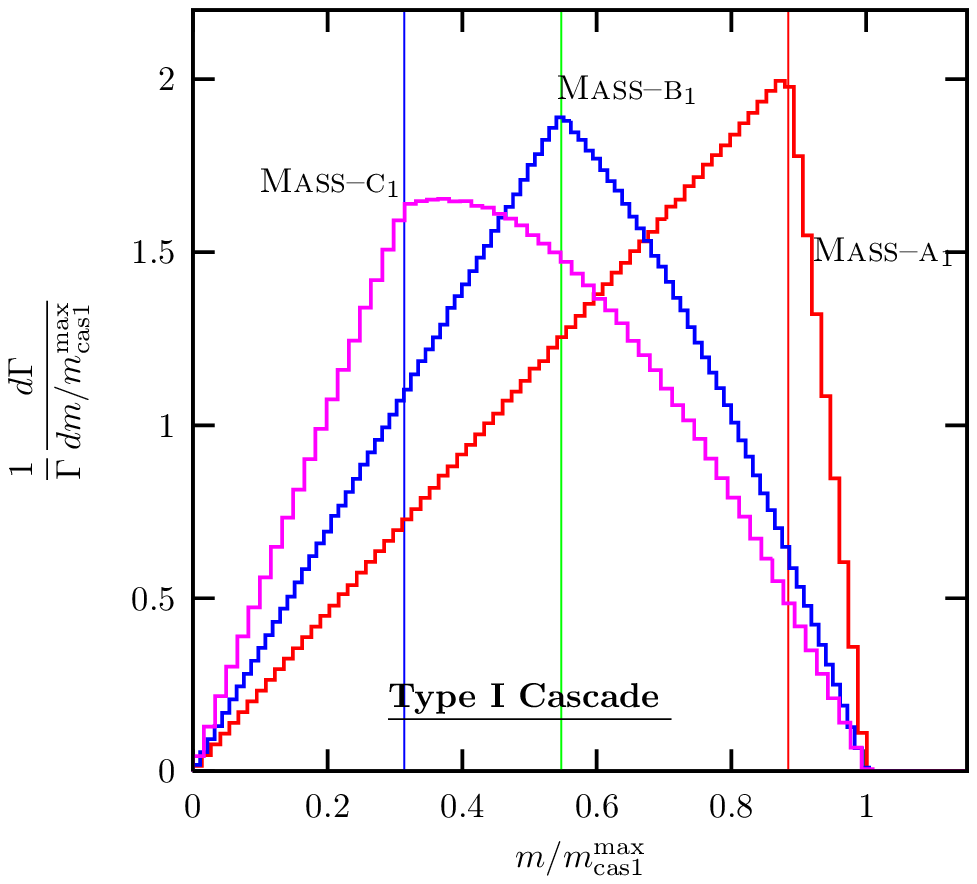}
  \caption{
The normalized differential decay rate of the invariant mass of two visible particles,
$\frac{d\Gm}{\Gm d m}$ for the \tyo~cascade decay.
The masses are in Table 
\protect\ref{table:cascade1:mass}.
}
\label{fig:cas1:m}
\end{figure}

In Fig.~\ref{fig:cas1:m},
we show the normalized differential decay rate of
$d\Gm/d m$.
In order to compare the cusp shapes only,
we present it as a function of $m/ \mmaxcaso$.
The vertical lines denote the positions of $\mcuspcaso$
in units of $\mmaxcaso$.
The \mao\;case with $\xi_B = 0.12$ has a very sharp $m$ 
cusp.
The \mbo\;case with $\xi_B =0.60$
shows a  triangular shape with a cusped peak.
However,
the \mco\;case with $\xi_B=1.16$ has a dull cusp.
In the $d \Gamma/dm$ distribution, the profile shape near the endpoint can 
be generally regarded as fast-dropping, as suggested by Eq. (\ref{eq:cas1:cusp:eq}).

\begin{figure}[!t]
\centering
\includegraphics[scale=0.8]{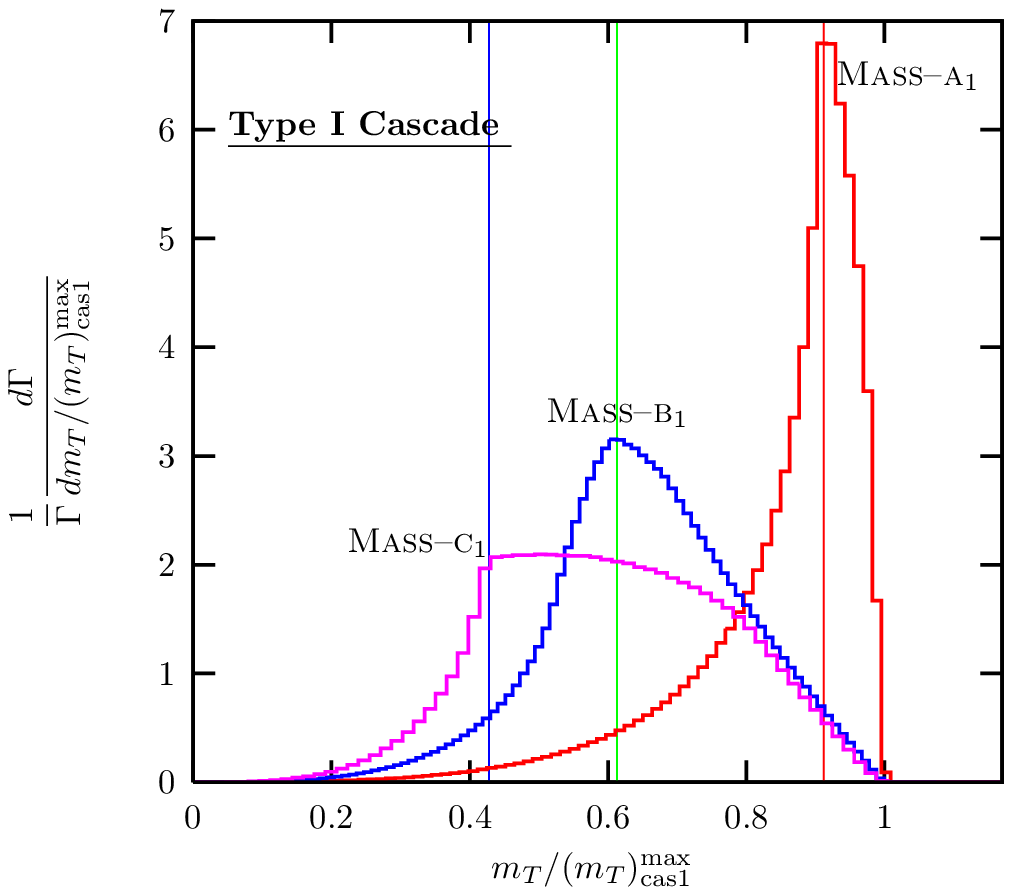}
\caption{
The normalized differential decay rate of the transverse mass of two visible particles,
$\frac{d\Gm}{\Gm d m_T}$  for the \tyo~cascade decay.
The masses are in Table \protect\ref{table:cascade1:mass}.}
\label{fig:cas1:mt}
\end{figure}

\vskip 0.5mm
\noindent
\underline{\textit{(ii) Transverse mass $m_T$ distribution}:}
Figure \ref{fig:cas1:mt} shows the rate of the transverse mass $m_T$ distributions. 
For all three \mao, \mbo\;and \mco\;cases, 
the $m_T$ distributions show visible cusp structures.
It is interesting to note that the \mco\;case has a more visible 
$m_T$ cusp compared with the $m$ cusp. 
We also note that
this is contrasted to the antler decay case where there is no cusp
in the $m_T$ distribution~\cite{long:antler}. 
As shall be shown in the next section,
the \tyt~cascade decay also has a cusp in the $m_T$ 
distribution. 
Therefore, the presence of the $m_T$ cusp 
can be used for identifying 
the cascade decay topology.
The cusp and maximum positions in terms of the masses are
\bea
\label{eq:cas1:mt:cusp:max}
\mtcuspcaso= E_n + E_f e^{\xi_C-\xi_B},
\quad
\mtmaxcaso = E_n + E_f e^{\xi_C+\xi_B},
\eea 
where $E_n$ and $E_f$ are in Eq.~(\ref{eq:En:Ef}).

\begin{figure}[!t]
\centering
  \includegraphics[scale=0.8]{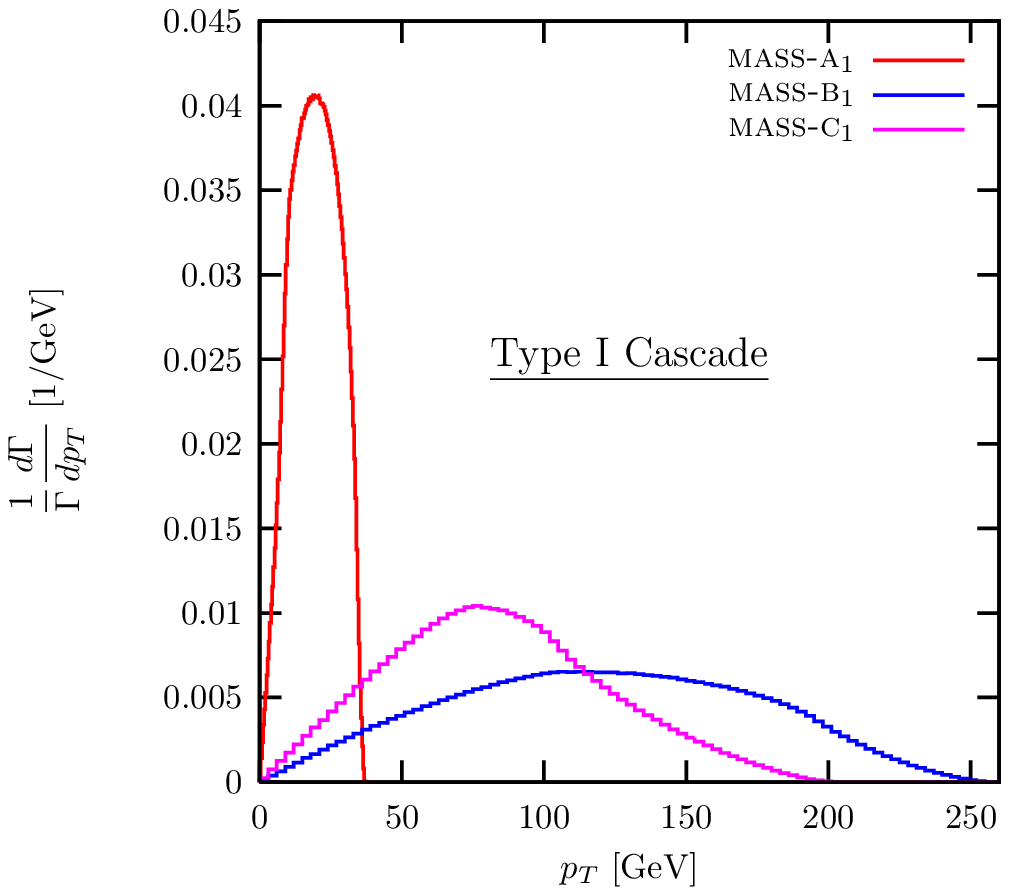}
  \caption
{
The normalized differential decay rate of the transverse momentum of two visible particles,
$\frac{d\Gm}{\Gm d p_T}$  for the \tyo~cascade decay
for the masses in Table \protect\ref{table:cascade1:mass}.}
\label{fig:cas1:pT}
\end{figure}

\vskip 0.5mm
\noindent
\underline{\textit{(iii) The system $p_T$ distribution}:}
Figure \ref{fig:cas1:pT}
shows the normalized distribution $d \Gm/d p_T$ 
of the transverse momentum $p_T$ of 
two visible particles.  
For all three mass spectra in Table \ref{table:cascade1:mass},
the $p_T$ distribution has smooth peak
without a cusp structure.
Still the endpoint of $p_T$ distribution can be observed,
which is 
\bea
\label{eq:cas1:pt:cusp:max}
\ptmaxcaso &=&
E_n + E_f e^{-\xi_C+\xi_B}.
\eea
Only the \mao\, case has a fast dropping endpoint shape,
which is attributed to very small momentum transfer to the visible particles.
More general cases of \mbo\;and \mco\;have long-tailed endpoints.
The $p_T$ distribution is not useful for the mass measurement.

\begin{figure}[!t]
\centering
  \includegraphics[scale=0.75]{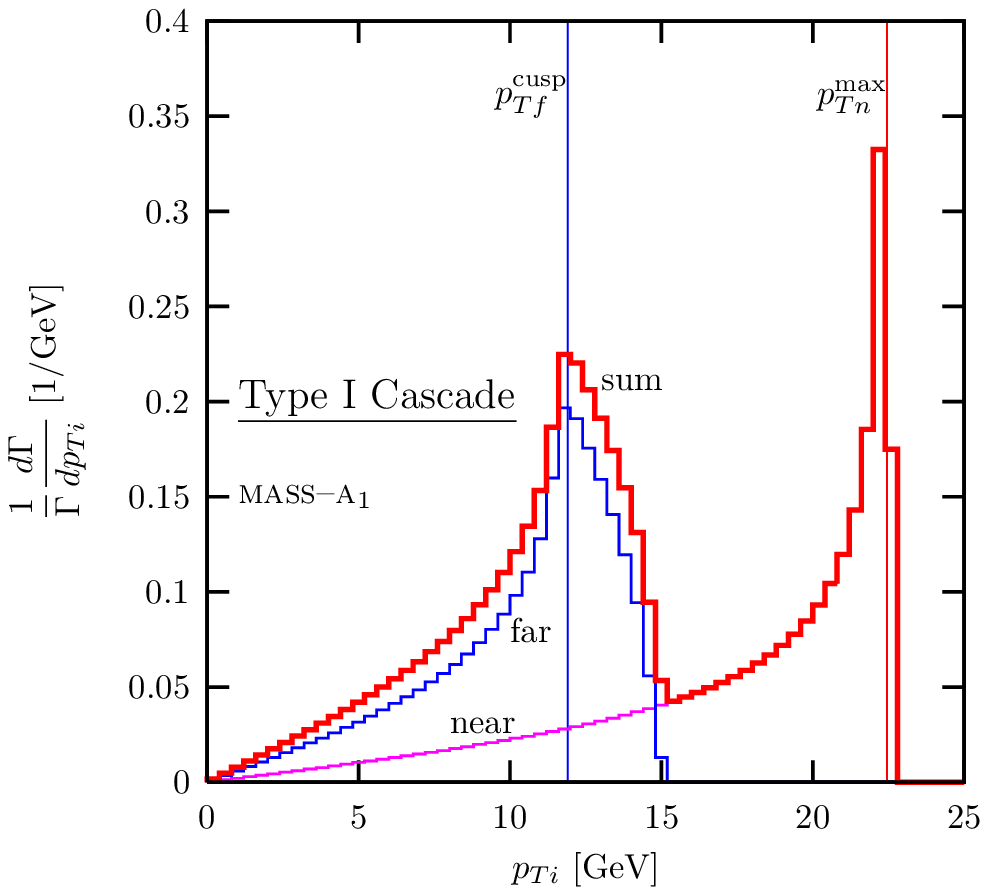}
  \phantom{x}
  \includegraphics[scale=0.75]{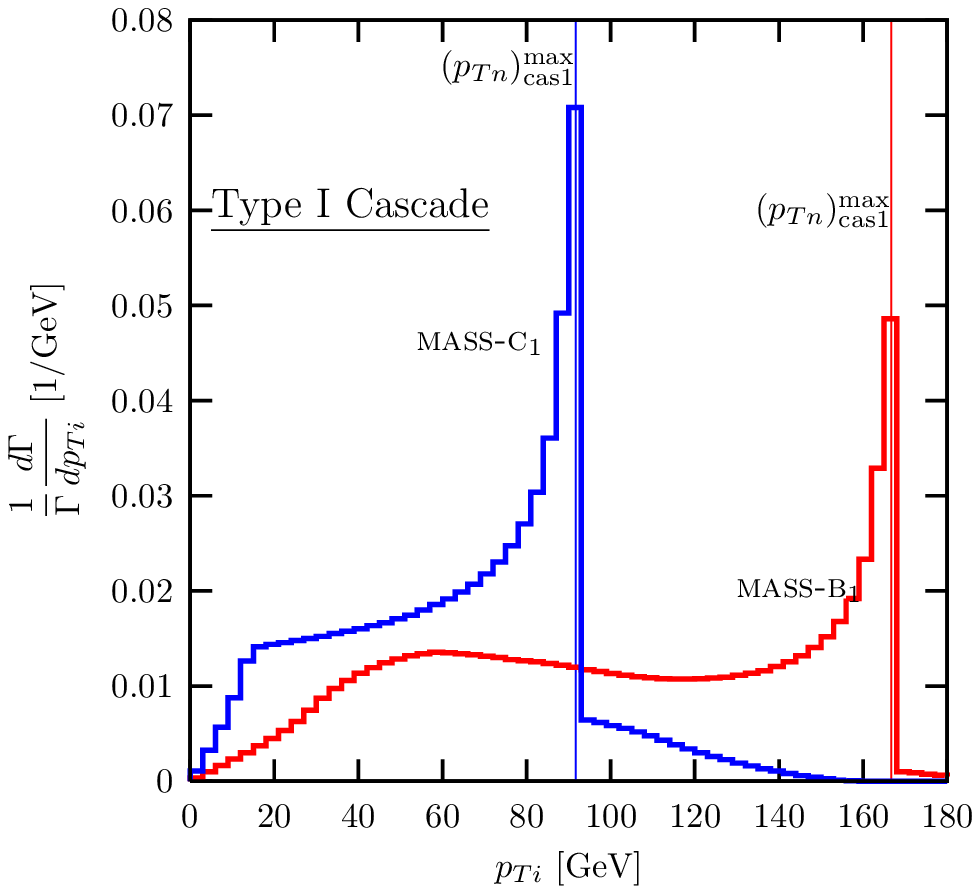}
  \caption{
The normalized differential decay rate of the transverse momentum of one visible particle,
$\frac{d \Gm}{\Gm d p_{Ti} }$ for the \tyo~cascade decay.
In the \mao\;case,
the line labeled by ``near" (``far") denotes the $p_{Ti}$ distribution of $a_1$
($a_2$).
Thick lines are the summed distributions of $p_{Ti}$.
}
\label{fig:cas1:pti}
\end{figure}

\vskip 0.5mm
\noindent
\underline{\textit{(iv) Single particle $p_{Ti}$ distribution}:}
The individual transverse momentum $p_{Ti}$ distributions of two 
visible particles show unique functional behaviors, 
as shown in Fig.~\ref{fig:cas1:pti}.
The thin solid line labeled by ``near" 
(``far") is the $p_{Ti}$ distribution
of the near visible particle $a_1$ (the far visible particle $a_2$).
The $p_{Tf}$ distribution
has both the cusp and the endpoint structures,
while the $p_{Tn}$ distribution has only an endpoint.
This $p_{Tn}$ endpoint has a sudden ending shape like a step function,
which holds true for all mass cases.

In most realistic situations, 
one may not distinguish the near 
visible particle from the far one.
Here we show a more practical observable,
the transverse momentum of any visible particle, which 
becomes the sum of both $p_{Ti}$ distributions.
The thick lines in Fig.~\ref{fig:cas1:pti} represent the sum

The position of the cusp and the endpoint in the $p_{Tf}$ distribution 
is given by 
\begin{eqnarray}
\left(p_{Tf}\right)^{\rm cusp}_{\rm cas1} = E_f e^{\xi_C - \xi_B }, \quad
\left(p_{Tf}\right)^{\rm max}_{\rm cas1} = E_f e^{\xi_C + \xi_B },
\label{eq:cas1:ptf:cusp:max}
\end{eqnarray}
and the endpoint in the $p_{Tn}$ distribution is located at
\begin{eqnarray}
\left(p_{Tn}\right)^{\rm max}_{\rm cas1} = E_n,
\label{eq:cas1:ptn:max}
\end{eqnarray}
where $E_n$, $E_f$, $\xi_B$ and $\xi_C$ are in Eq. (\ref{eq:cas1:rapidity})
and (\ref{eq:En:Ef}). Depending on whether 
$\left(p_{Tn}\right)^{\rm max}_{\rm cas1} > \left(p_{Tf}\right)^{\rm max}_{\rm cas1}$  
(the case of \mao) or not (the cases of \mbo\;and \mco), 
the summed distribution shows apparently different shape, as
 shown in Fig.~\ref{fig:cas1:pti}. While $\left(p_{Tn}\right)^{\rm max}_{\rm cas1}$
can be easily determined due to the unique spiky feature of the distribution, the cusp 
and the endpoint of the $p_{Tf}$ distribution when 
$\left(p_{Tn}\right)^{\rm max}_{\rm cas1} < \left(p_{Tf}\right)^{\rm max}_{\rm cas1}$ is 
rather difficult to identify in the summed $p_{Ti}$ distribution. 

\begin{figure}[!t]
\centering
  \includegraphics[scale=0.8]{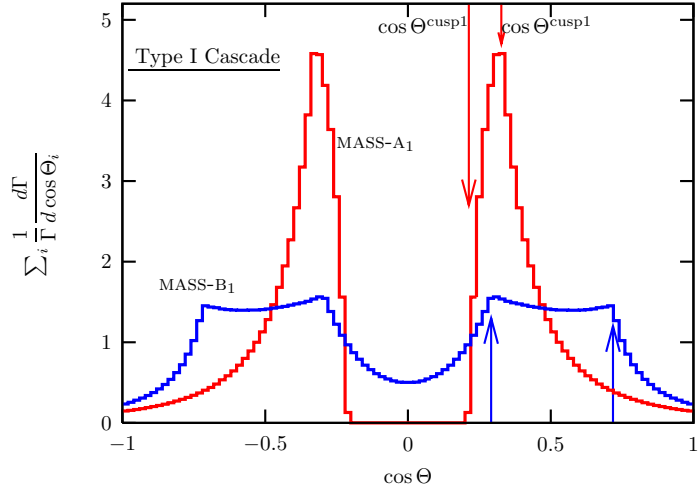}
  \caption{
The summed distributions of
$\cos\Theta_i$ in the \tyo~cascade decay
for the masses in Table \ref{table:cascade1:mass}.}
\label{fig:cas1:costh}
\end{figure}

\vskip 0.5mm
\noindent
\underline{\textit{(v) $\cos\Theta$ distribution}:}
The variable $\cos\Th$ in Eq.~(\ref{eq:cos:def})
is defined by the angle of 
\emph{one} visible particle.
We have two $\cos\Th$ distributions for $a_1$ and $a_2$,
which cannot be distinguished.
In Fig.~\ref{fig:cas1:costh},
therefore,
we present the summation of two $\cos\Th_i$ distributions
in the rest frame of $D$
for \mao\;and \mbo\;cases.
It is symmetric about $\cos\Theta =0$,
and has two cusp structures,  
$\cos\Theta^{\rm cusp1}_{\rm cas1}$ and 
$\cos\Theta^{\rm cusp2}_{\rm cas1}$,
marked by the vertical arrows. 
In terms of masses, they are
\bea
\label{eq:cas1:costh}
\cos\Theta^{\rm cusp1}_{\rm cas1} =
\frac{E_n -E_f \exp(\xi_B - \xi_C )}
{E_n +E_f \exp(\xi_B - \xi_C )}
,
    \quad    \cos\Theta^{\rm cusp2}_{\rm cas1} =
\frac{E_n
    - E_f \exp(-\xi_B-\xi_C)}
    {E_n
    + E_f \exp(-\xi_B-\xi_C)}.
\eea
In the \mao\;case,  $\cos\Theta^{\rm cusp1}_{\rm cas1} $ stands
on  a steep slope, which is difficult to probe.
The \mbo\;case shows two pronounced cusps.

\section{\tyt~cascade decay}
\label{sec:type2}

\tyt~cascade decay is a chain decay of
\bea
D(P) \longrightarrow  && C + X_1,       \\
\no
          && C \longrightarrow       B + a_1(k_1) ,  \\ \no
          &&  \hspace{30pt}
 B \longrightarrow      a_2(k_2) + X_2.
\eea
A good example can be found in the MSSM:
\bea H/A \to \neuo + \neu_2, \quad \neu_2 \to
\ell_n + \tilde{\ell}, \quad \tilde{\ell} \to \ell_f + \neuo. 
\eea

\begin{table}[t!]
\centering
\begin{tabular}{|c||c|c|c|c||c|}
\hline
         & ~~$m_D$~~ & ~~$m_C$~~ & ~~$m_B$~~&~~$m_X$~~
            &  ~~$m^{\rm max}$~~  \\
\hline
~~~{\sc Mass--a}$_2$~~~  & 614  & 299 & 222 & 161 & 138.0  \\
~~~{\sc Mass--b}$_2$~~~  & 600 & 300 & 200 & 100 & 193.6 \\
~~~{\sc Mass--c}$_2$~~~  & 400 & 250 & 150 & 120 & 120.0 \\
\hline
\end{tabular}
\caption{
Test mass spectrum sets for the
\tyt~cascade decay. All masses are in units of GeV. }
\label{table:cascade2:mass} 
\end{table}

As in the \tyo~cascade decay,
we restrict ourselves to the realistic cascade decay with
$m_{a_1}=m_{a_2}=0$ and $m_{X_1}=m_{X_2}$.
Then there
are two independent rapidities, 
$\zeta_C$ and $\zeta_B$: 
\bea
\cosh\zeta_B = \frac{m_C}{2 m_B} \left( 1 + \frac{m_B^2}{m_C^2}
\right),
\quad
\cosh\zeta_C = \frac{m_D}{2 m_C} \left( 1+ \frac{m_C^2}{m_D^2}
-\frac{m_X^2}{m_D^2} \right). 
\eea
For illustration, we take three mass
sets for the \tyt~cascade decay
in Table \ref{table:cascade2:mass}.

\begin{figure}[!t]
\centering
  \includegraphics[scale=0.8]{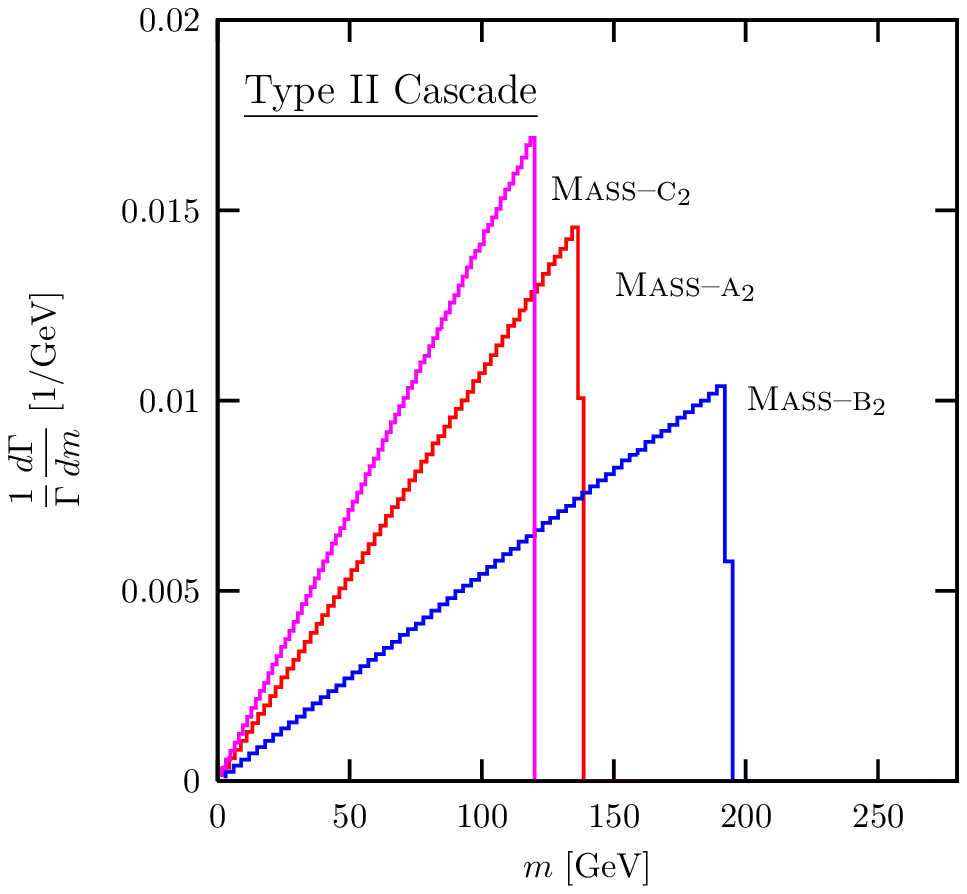}
  \caption
{
The normalized differential decay rate of the invariant mass of two visible particles,
$\frac{d\Gm}{\Gm d m}$
for the \tyt~cascade decay. 
The mass spectrum sets are
described in Table \protect\ref{table:cascade2:mass}.}
\label{fig:cas2:m}
\end{figure}

\vskip 0.5mm
\noindent
\underline{\textit{(i) Invariant mass $m$ distribution}:}
We first study the distribution of the invariant mass of
$a_1$ and $a_2$.
Note that in the view point of $a_1$ and $a_2$,
this \tyt~cascade decay is a three body decay of the parent particle $C$.
The presence of the invisible $X_1$ decayed from $D$
does not
change any Lorentz invariant result.
The $m$ distribution is the same as that of, \textit{i.e.},
$m_{\ell\ell}$ of the decay $\neu_2 \to \ell_n
\tilde{\ell} \to \ell_n\ell_f \neuo$ in the MSSM.
This $m_{\ell\ell}$ distribution is well known to have no
cusp structure.
The endpoint is \cite{susy:mll}
\bea 
\left( \mmaxcast
\right)^2 = m_C^2 \left(1 - \frac{m_B^2}{m_C^2} \right) \left(1 -
\frac{m_X^2}{m_B^2} \right).
\eea 
In Fig.~\ref{fig:cas2:m}, we show the $m$ distribution
for three sets of the mass parameters in Table
\ref{table:cascade2:mass},
all of which have right-angled triangle shoe without a cusp.

The absence of a cusp in a two-step cascade decay can be understood
by a simple kinematic configuration.
For the antler decay ($D \to B_1 + B_2 \to a_1 X_1 + a_2 X_2$)
in the massless visible particle case ($m_{a_1} = m_{a_2} =0$),
the following four critical points 
correspond to a kinematic singular structure~\cite{long:antler}:
\bea
{\renewcommand{\arraystretch}{1.2}
\begin{array}{c|c c}
\hbox{1D configuration } & m_{a_1 a_2} \\ \hline
~~~\stackrel{a_2}{\Longleftarrow}
~~\stackrel{B_2}{\longleftarrow}  
~~\stackrel{D}{\bullet}~~ 
\stackrel{B_1}{\longrightarrow} 
~~
\stackrel{a_1}{\Longrightarrow}~~~ 
& ~~~\hbox{ max }~~~ \\
\stackrel{a_2}{\Longrightarrow}
~~\stackrel{B_2}{\longleftarrow}  
~~\stackrel{D}{\bullet}~~ 
\stackrel{B_1}{\longrightarrow} 
~~
\stackrel{a_1}{\Longleftarrow} 
& \hbox{ cusp } \\
\stackrel{a_2}{\Longrightarrow}
~~\stackrel{B_2}{\longleftarrow}  
~~\stackrel{D}{\bullet}~~ 
\stackrel{B_1}{\longrightarrow} 
~~
\stackrel{a_1}{\Longrightarrow} 
& \hbox{ min } \\
\stackrel{a_2}{\Longleftarrow}
~~\stackrel{B_2}{\longleftarrow}  
~~\stackrel{D}{\bullet}~~ 
\stackrel{B_1}{\longrightarrow} 
~~
\stackrel{a_1}{\Longleftarrow} 
& \hbox{ min } \\
\end{array}
}
\eea
Here we simplify the picture as an one-dimensional case.
It is clear to see that
$m_{a_1 a_2}^{\rm min}$ 
happens when two observable particles move
in the same direction,
while two kinematic configurations of back-to-back moving 
correspond to either $m_{a_1 a_2}^{\rm max}$ or $m_{a_1 a_2}^{\rm cusp}$. 
For a two-step cascade decay ($C \to a_1 +B \to a_1 + a_2 X_2$),
$a_1$ and $a_2$ in one-dimensional space have
only two independent kinematic configurations,
moving in the same direction and moving in the opposite direction.
There is no critical point left for the cusp.

\begin{figure}[!t]
\centering
  \includegraphics[scale=0.8]{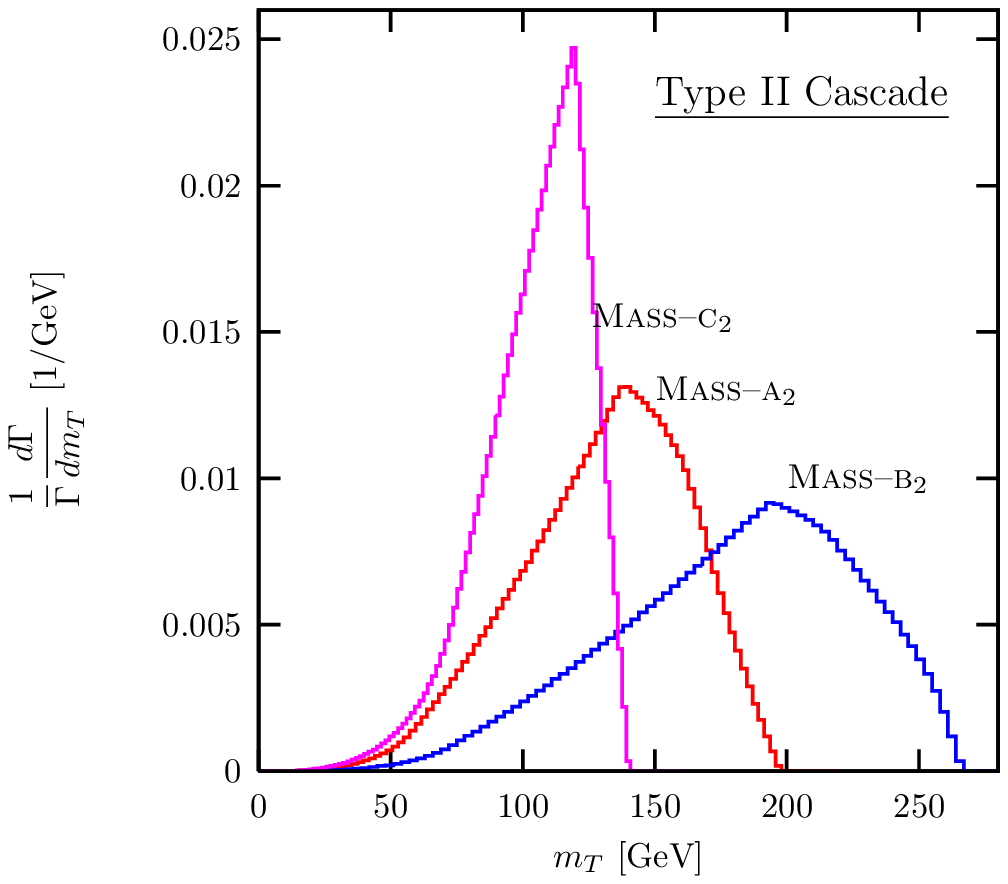}
  \caption
{
The normalized differential decay rate of the transverse mass of
two visible particles, $\frac{d\Gm}{d m_T}$ for the \tyt~cascade decay. 
The mass spectrum sets are
described in Table \protect\ref{table:cascade2:mass}.}
\label{fig:cas2:mT}
\end{figure}

\vskip 0.5mm
\noindent
\underline{\textit{(ii) Transverse mass $m_T$ distribution}:}
Unlike the invariant mass distribution,
the $m_T$ distribution
contains the information about the 
transverse momenta of the first missing particle $X_1$.
As shown in Fig.~\ref{fig:cas2:mT},
there is a cusp here.
We stress once again that this $m_T$ cusp appears only when 
$D$ is produced at rest in the transverse direction.

Another interesting feature is that the position of the $m_T$ cusp is
nothing but the $m$ maximum:
\bea
\mtcuspcast =\mmaxcast.
\eea 
This non-trivial equality is a unique feature of
the \tyt~cascade decay.

\begin{figure}[!t]
\centering
   \includegraphics[scale=0.8]{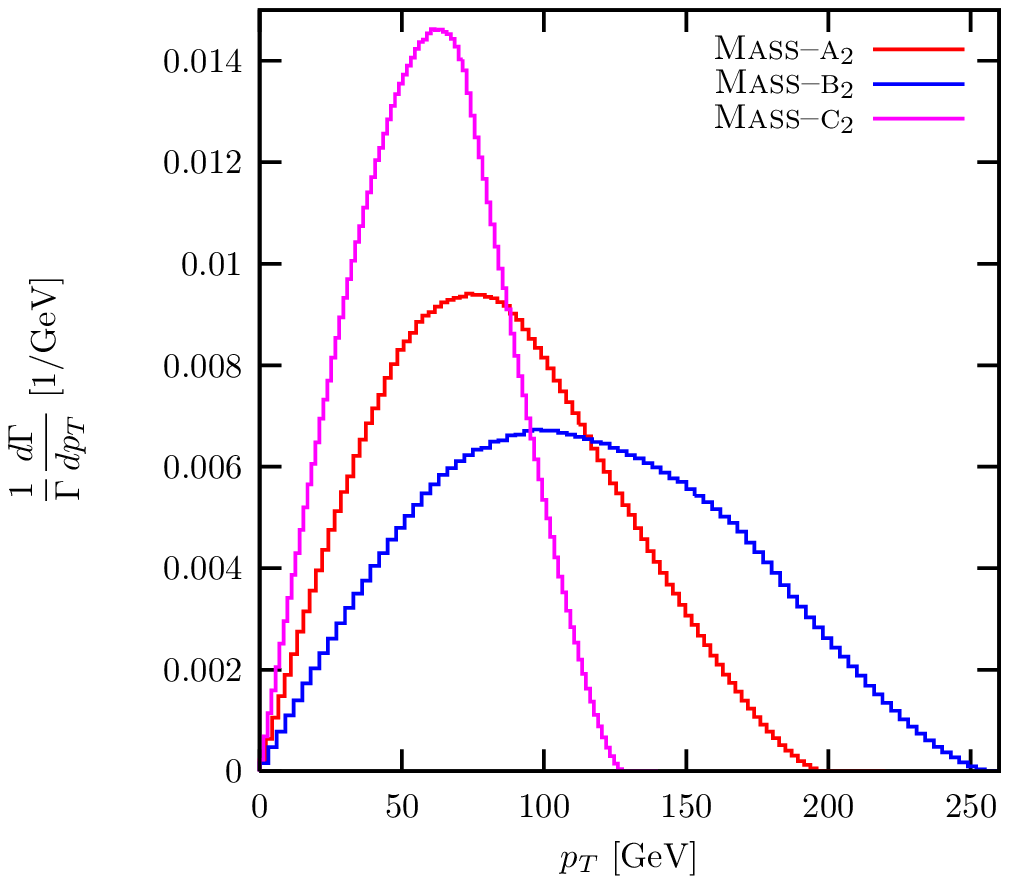}
  \caption
{
The normalized differential decay rate of the transverse momentum of two visible particles,
$\frac{d\Gm}{\Gm d p_T}$ for the
\tyt~cascade decay. The mass spectrum sets are in
Table \protect\ref{table:cascade2:mass}.}
\label{fig:cas2:pT}
\end{figure}

\vskip 0.5mm
\noindent
\underline{\textit{(iii) System $p_T$ distribution}:}
The total $p_T$ distributions for the \tyt~cascade decay
are shown
in Fig.~\ref{fig:cas2:pT}.
All three mass sets  
have smooth $p_T$ distributions.
And their endpoints are all long-tailed.
This feature is common for 
the antler, \tyo, and \tyt~cascade decay topology.

\begin{figure}[!t]
\centering
  \includegraphics[scale=0.75]{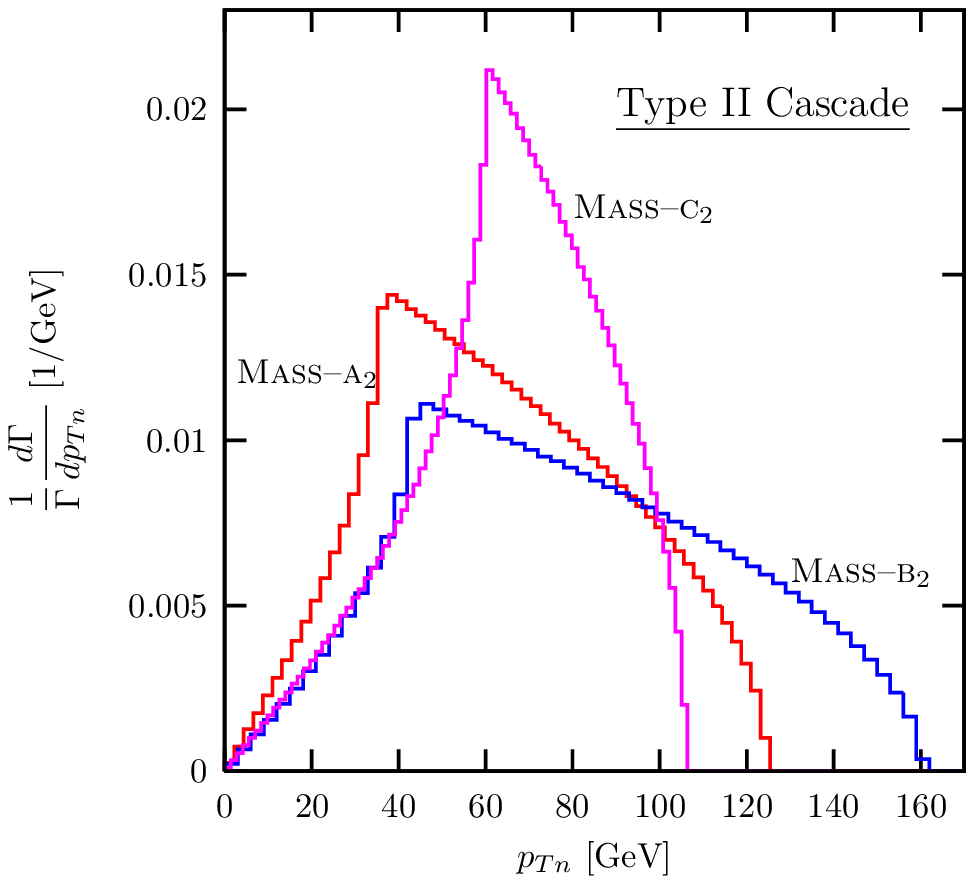}
  \phantom{x}
  \includegraphics[scale=0.75]{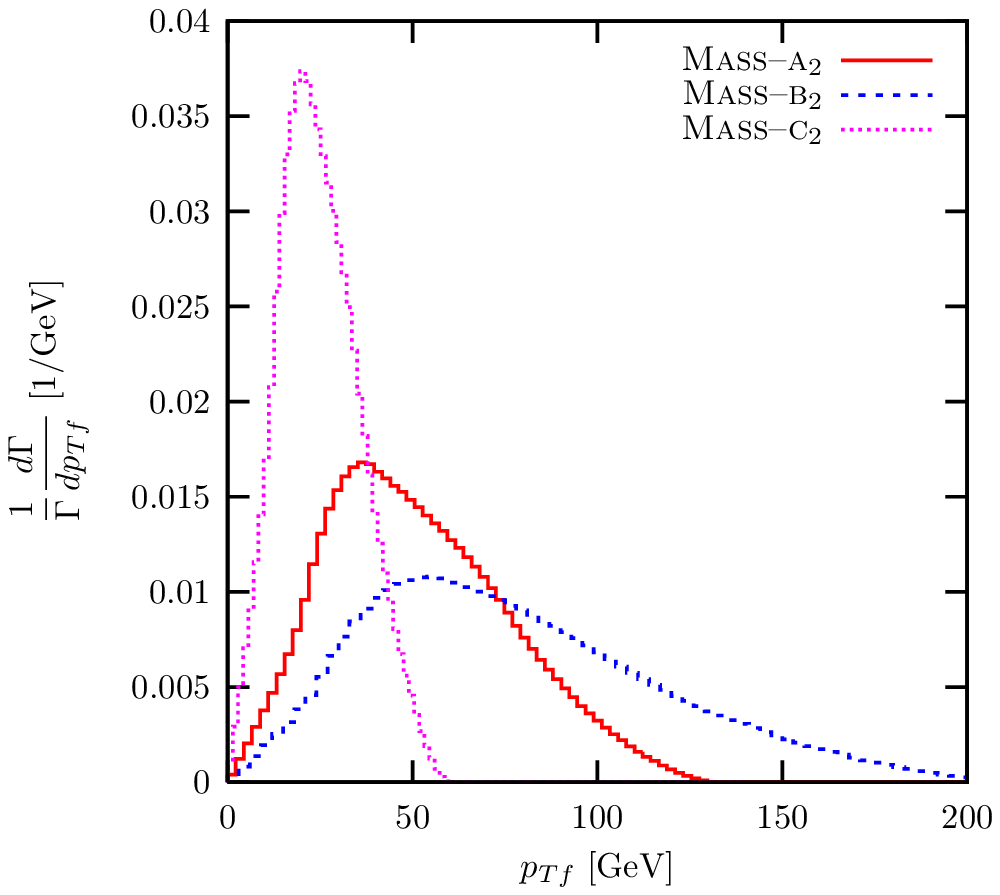}
  \caption[The normalized differential decay rate of $d\Gm/d m$ for the
massless SM particles.]
{
The normalized differential decay rate of the transverse mass
of one visible particle, $\frac{d\Gm}{\Gm d {p_{Ti}}}$ for the
\tyt~cascade decay.
The left figure is for the near visible particle $a_1$,
and the right one is for the far visible particle $a_2$.}
\label{fig:cas2:pTnf}
\end{figure}

\vskip 0.5mm
\noindent
\underline{\textit{(iv) Single particle  $p_{Ti}$ distribution}:}
Figure \ref{fig:cas2:pTnf} 
shows the distribution of
the individual transverse momentum of the near $a_1$ and
the far $a_2$.
The near $p_{Tn}$ distribution has 
a sharp cusp 
and a fast dropping endpoint.
However the $p_{Tf}$ distribution has a long tailed endpoint without any cusp.
In terms of masses they are simply
\bea
\ptncuspcast
= \frac{m_C}{2} \left( 1-\frac{m_B^2}{m_C^2}
\right) e^{-\zeta_C}, \quad \ptnmaxcast = \frac{m_C}{2} 
\left( 1-\frac{m_B^2}{m_C^2} \right)
e^{\zeta_C}. 
\eea 
Note that the product of $\ptncuspcast$ and $\ptnmaxcast$
removes the $\zeta_C$ dependence, which depends on the
intermediate masses $m_C$ and $m_B$. 
In addition the ratio $\ptncuspcast / \ptnmaxcast$ depends only on the
rapidity $\zeta_C$.

\begin{figure}
\centering
  \includegraphics[scale=0.8]{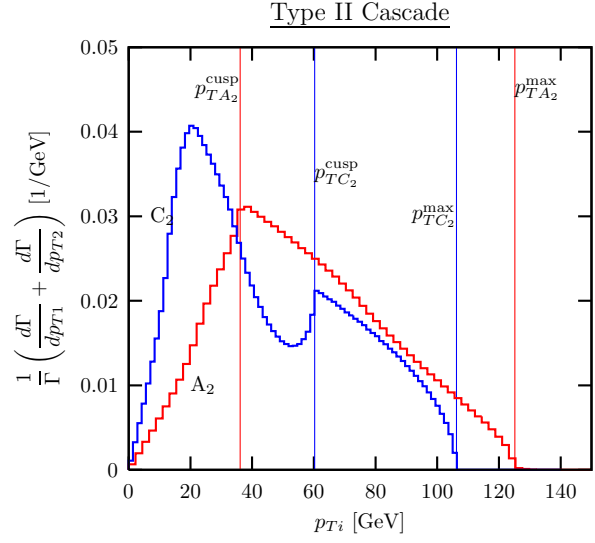}
  \caption{
The sum of two normalized differential decay rate with 
respect to the individual transverse momenta
of the near and far visible particles.}
\label{fig:cas2:pt:sum}
\end{figure}

As discussed before, the individual $p_{Ti}$ distribution
cannot be constructed.
Instead we show the sum of two distributions in Fig.~\ref{fig:cas2:pt:sum}.
For the \mat\;case, the 
cusp in the $p_{Tn}$ distribution
and the smooth peak of the $p_{Tf}$ distribution are located nearby.
In their sum, the $p_{Tn}$ 
cusp survives over the relatively round $p_{Tf}$ peak 
and the fast dropping $p_{Tn}$ endpoint is also
measurable.
For the \mct\;case,
however,
the $p_{Tn}$ cusp and the $p_{Tf}$ peak
are separated so that the summed distribution
shows both.
With finite number of data,
it would be difficult to distinguish the $p_{Tn}$ 
cusp from the $p_{Tf}$ peak.

\begin{figure}[!t]
\centering
  \includegraphics[scale=0.8]{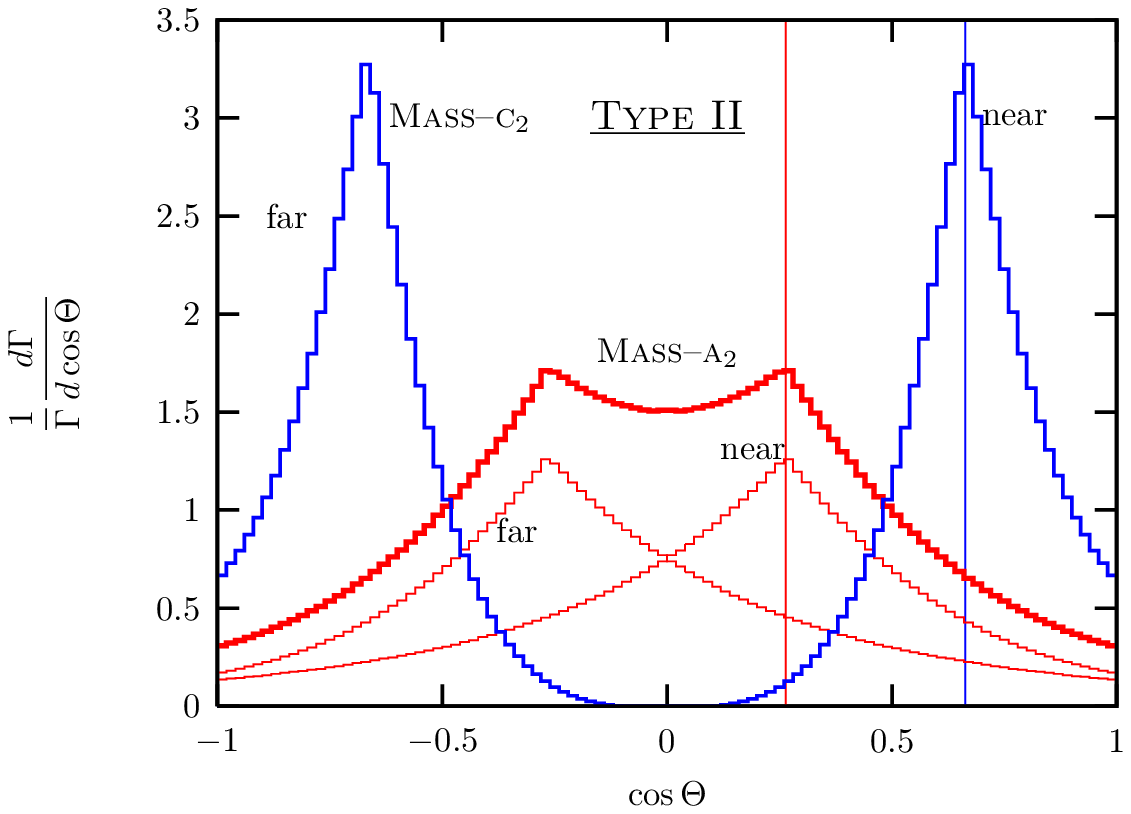}
  \caption{
The sum of $d\Gm/d \cos\Theta_i$ for the $\tyt$~cascade decay.}
\label{fig:cas2:costh}
\end{figure}

\vskip 0.5mm
\noindent
\underline{\textit{(v) $\cos\Theta$ distribution}:}
We consider the $\cos\Theta$ distribution 
for the \tyt~cascade decay.
In Fig.~\ref{fig:cas2:costh} 
we show 
the normalized $d \Gm /d \cos\Theta$
for the near and far visible particles
(denoted by thin lines) as well as their sum (thick lines)
for the \textsc{Mass--A}$_2$ and
\textsc{Mass--C}$_2$. 
In both cases, the 
summed distribution of $\cos\Th_i$ is symmetric about $\cos\Th=0$,
and has one independent sharp cusp 
denoted by vertical lines in Fig.~\ref{fig:cas2:costh}. 
The $\cos\Th$ cusp position in terms of the mass parameters
is 
\bea 
\cos\Theta^{\rm cusp}_{\rm cas2} =
\frac{m_C\left(1-\frac{m_B^2}{m_C^2}\right)
         - m_B\left(1-\frac{m_X^2}{m_B^2}\right) e^{-\zeta_B}}
{m_C\left(1-\frac{m_B^2}{m_C^2}\right)
         + m_B\left(1-\frac{m_X^2}{m_B^2}\right) e^{-\zeta_B}}.
\eea

\vskip 0.5mm
\noindent
\underline{\textit{(vi) Mass determination from the cusps and endpoints}:}
Unlike the antler decay with one kind of intermediate particles,
the cascade decay has two different intermediate particles.
In addition,
the \tyt~decay has fewer independent observables of cusps and endpoints.
First 
there is no $m$ cusp structure.
Second the $m_T$ cusp position is the same as the
$m$ endpoint.
A natural concern arises whether we have enough information to
determine all the masses, especially at the LHC where
the $\cos\Theta$ cusp
cannot be used.
Fortunately 
three unknown masses
($m_C$, $m_B$, and $m_X$) 
are unambiguously determined by 
three
observables of
$\mmaxcast$, $\ptncuspcast$, and $\ptnmaxcast$:
\bea
\label{app:Cas1:mass:determination} 
m_C &=& R_\al m_D, \quad m_B =
\sqrt{1 - \frac{\alpha_1}{R_\alpha}}\; m_C, 
\quad m_X = \sqrt{1 -
\frac{\alpha_2}{R_\alpha}} \; m_B, 
\eea 
where $R_\alpha$ is
\bea 
R_\alpha &=& \frac{1
+\al_1\al_2}{\al_3 -\al_1-\al_2},
\eea
and $\al_{1,2,3}$ are
\bea
\alpha_1 &=& \frac{(\mmaxcast)^2}{2 m_D \sqrt{{\ptnmaxcast}/{
\ptncuspcast}}},
\\ \no
\alpha_2 &=& \frac{2 \sqrt{\ptnmaxcast/ \ptncuspcast}}{m_D} ,
\\ \no
\alpha_3 &=& \sqrt{\frac{\ptnmaxcast}{\ptncuspcast}} +
\sqrt{\frac{\ptncuspcast}{\ptnmaxcast}}. 
\eea

\section{Effects of realistic considerations}
\label{sec:effects}

All the previous expressions of the cusps and endpoints
have been derived in an idealistic situation:
the total decay widths of decaying particles are ignored; 
the $D$ rest frame is assumed to be reconstructed;
the spin-correlation effects from the full matrix elements
are negligible.
In this section,
we investigate these effects 
on the position and shape of each kinematic cusp and endpoint.

\subsection{Finite width effects}
\label{subsec:width}

Up to now we have applied the narrow width approximation,
ignoring the width of decaying particles.
Since the effect of finite $\Gm_D$ is very minor~\cite{cusp},
we focus on the effects of $\Gm_B$ and $\Gm_C$.

\begin{figure}[!t]
\centering
\includegraphics[scale=0.75]{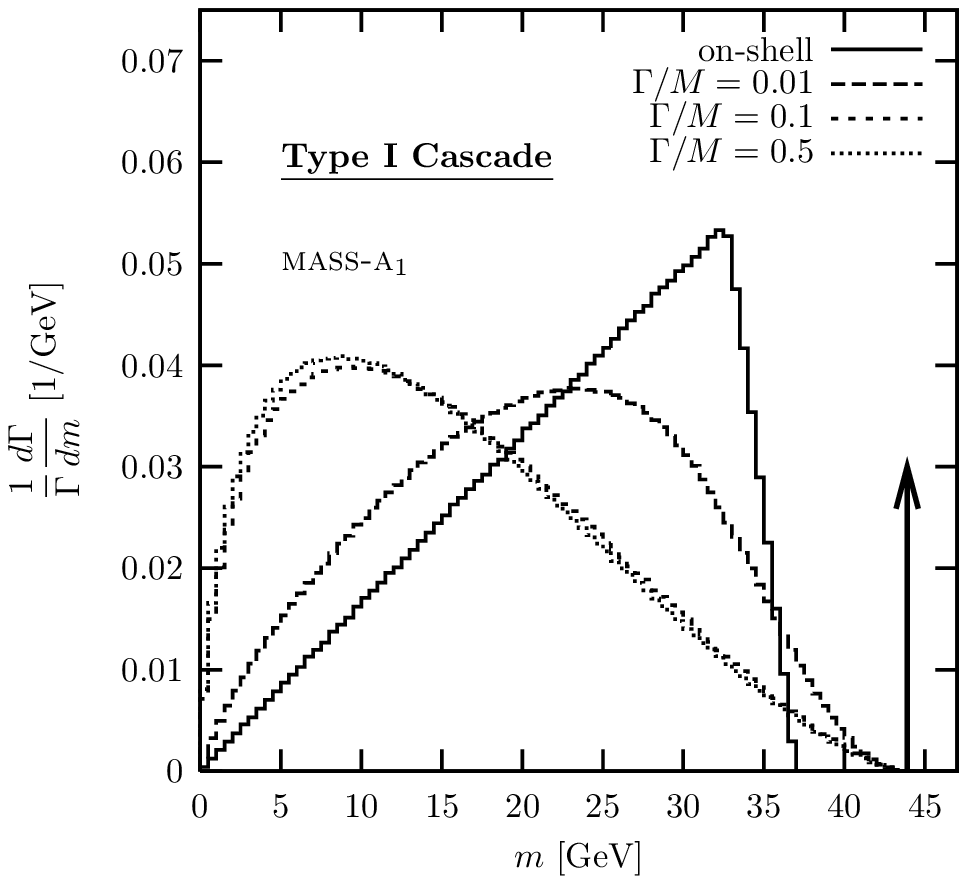}
\includegraphics[scale=0.75]{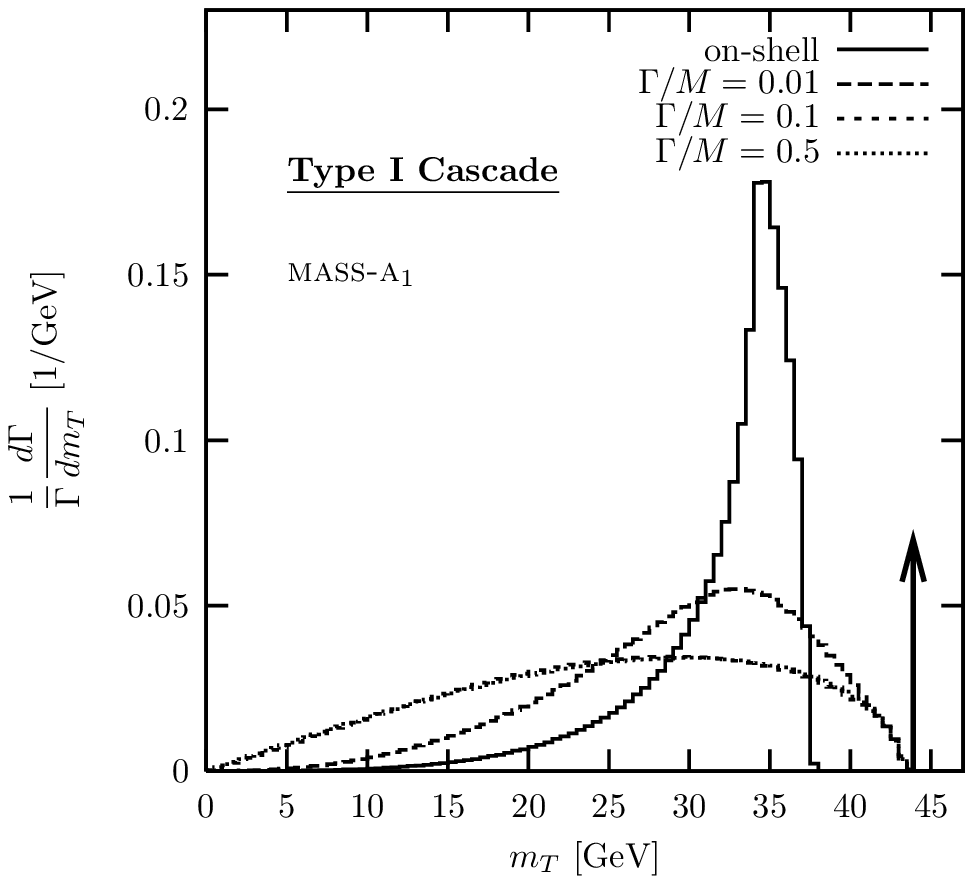}
  \caption{
The finite decay width effects on the $m$ and $m_T$
distributions in the \mao\;case.
Solid lines are for the on-shell decay,
the long dashed lines for 
$\Gm/M =0.01$,
the short dashed lines for $\Gm/M=0.1$,
and the dotted lines for $\Gm/M =0.5$.
Here $\Gm/M \equiv \Gm_B /M_B = \Gm_C /M_C$.
}
\label{fig:gm:m:mt:cas1A}
\end{figure}

\begin{figure}[!t]
\centering
\includegraphics[scale=0.75]{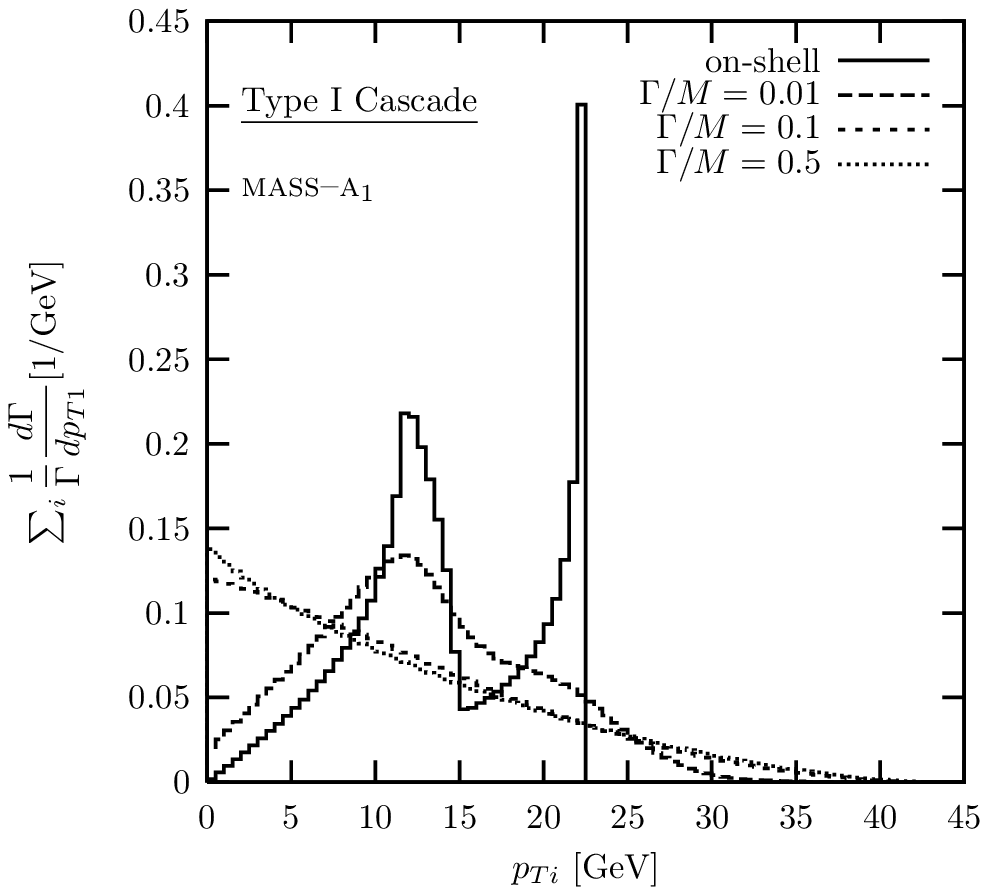}
\includegraphics[scale=0.75]{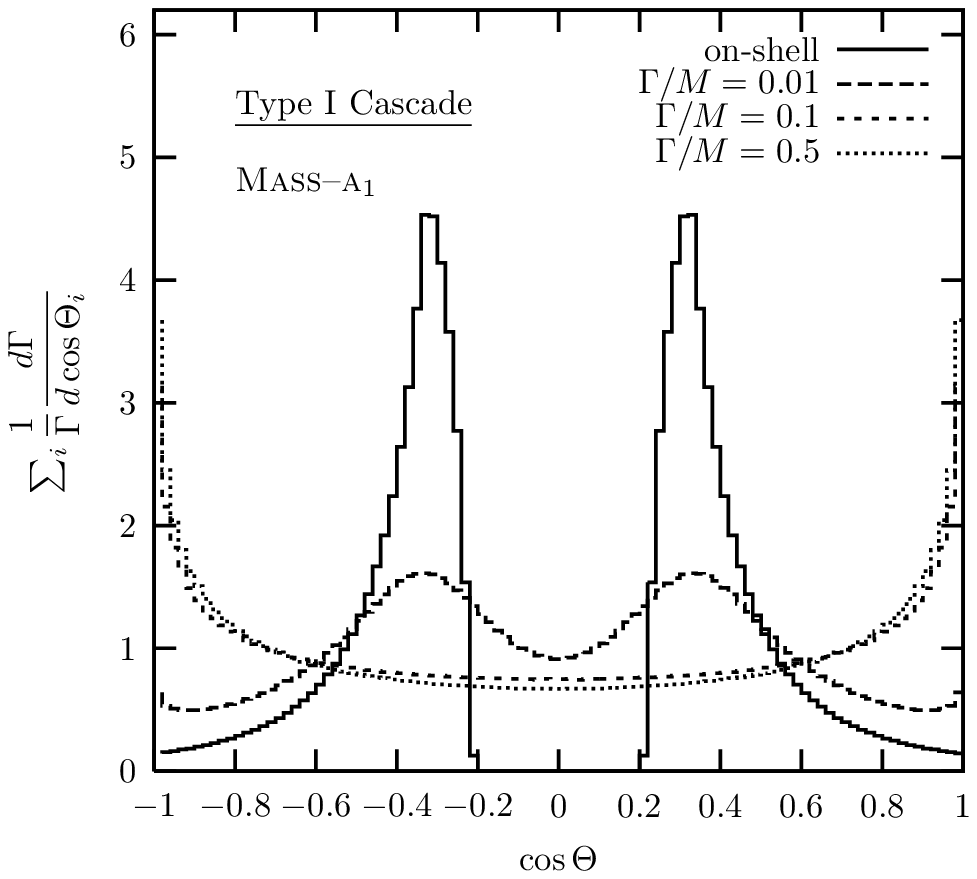}
  \caption{
The finite decay width effects on the summed distributions
of $p_{Ti}$ and $\cos\Theta_i$
 in the \mao\;case of the \tyo~cascade decay.
As before, we take
$\Gm/ M =0,~0.01,~0.1,~0.5$.}
\label{fig:gm:pti:costh:cas1}
\end{figure}

We find that the mass spectrum is the most crucial factor 
to determine the stability of the cusp and endpoint structures
under the width effects.
Out of six cases in Tables 
\ref{table:cascade1:mass} and \ref{table:cascade2:mass},
the \mao\;has very vulnerable structures.
This case is special because of its degenerate masses:
the observable particles have very
small momentum transfer
and their kinematic phase space is highly limited.

In Fig.~\ref{fig:gm:m:mt:cas1A} and Fig.~\ref{fig:gm:pti:costh:cas1},
we show the finite width effects for the \mao\;case.
We present four cases for $\Gm_B$ and $\Gm_C$:
on-shell (solid line),
$\Gm/M =0.01$ (long dashed line), $\Gm/M=0.1$ (short dashed lien),
and $\Gm/M =0.5$ (dotted line).
Here $\Gm/M \equiv \Gm_B/ m_B =\Gm_C/ m_C$ for simplicity.
Just one percent of $\Gm/M$ destroys all the sharp cusp structures
into smooth peaks.
In addition, the positions of the peaks
are shifted significantly from the true cusp positions.
There is no way to extract the mass information from the cusps.
For $\Gm/M \gsim 0.1$ the summed $p_{Ti}$ and $\cos\Theta_i$ distributions
lose their functional behaviors completely,
leaving very smooth and featureless distributions.

The fast-falling
endpoints in the $m$, $m_T$,
and $p_{Ti}$ distributions are also smeared out
considerably.
The degree of its shifting is 
large even for $\Gm/M =1\%$.
One interesting observation is that two shifted
endpoints of the $m$ and $m_T$ distributions
are the same to be $m_D-2 m_X$,
denoted by vertical arrows.
This new endpoint is from the kinematic
configuration where
two visible particles' momenta span 
all the phase space determined by
$m_D$ and $m_X$.
Even though we do not know the intermediate particle masses,
the missing particle mass $m_X$ can be read off.
For this information, the $m_T$ distribution is
more advantageous than the $m$ distribution,
because of its fast falling shape.

In a realistic new physics process,
however, 
this \mao\;case
does not allow even one percent of $\Gm/M$.
For example, the $\zt$ decay in the mUED model
has the decay widths of
\bea
\Gm_D = \Gm_\zt \simeq 270 \mev ,
\quad
\Gm_C = \Gm_\lt \simeq 5 \mev,
\quad
\Gm_B = \Gm_\lo \simeq 1 \mev,
\eea
which leads to $\Gm/M \sim 10^{-5}$.
This is attributed to the limited phase
space.
In summary, the extreme \mao\;case
has generically negligible width effects.
The cusp and endpoint structures are reserved.

\begin{figure}[!t]
\centering
\includegraphics[scale=0.75]{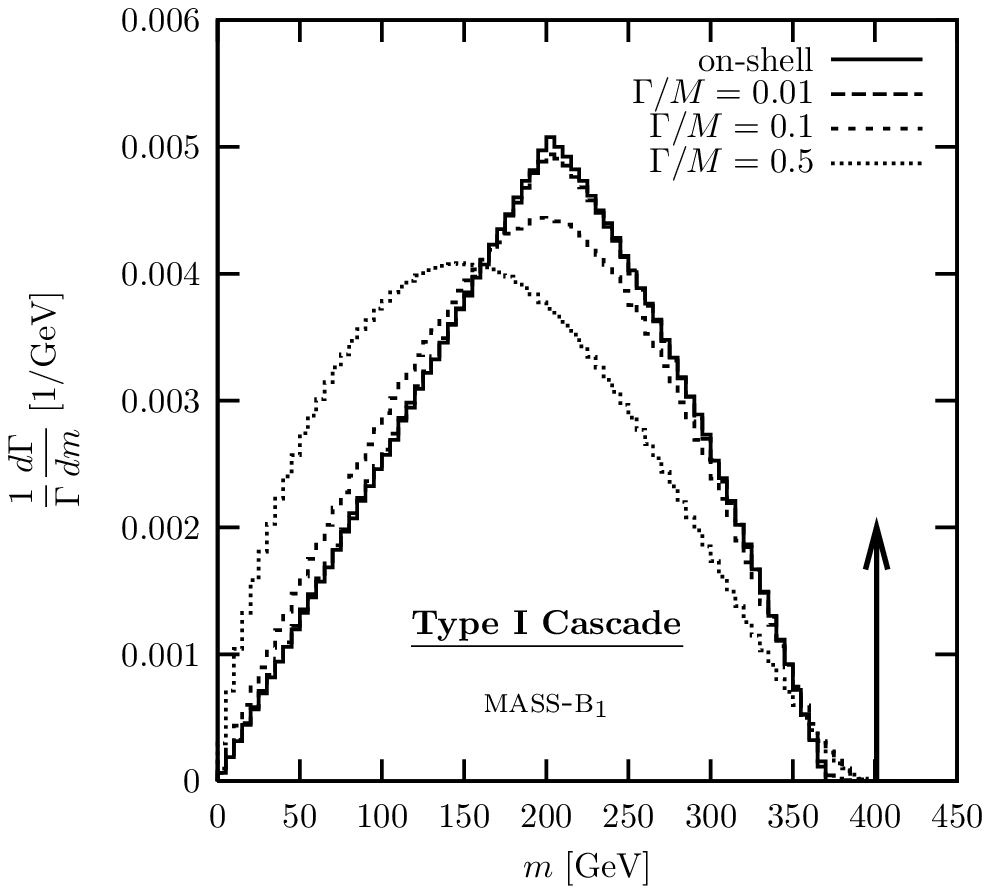}
\includegraphics[scale=0.75]{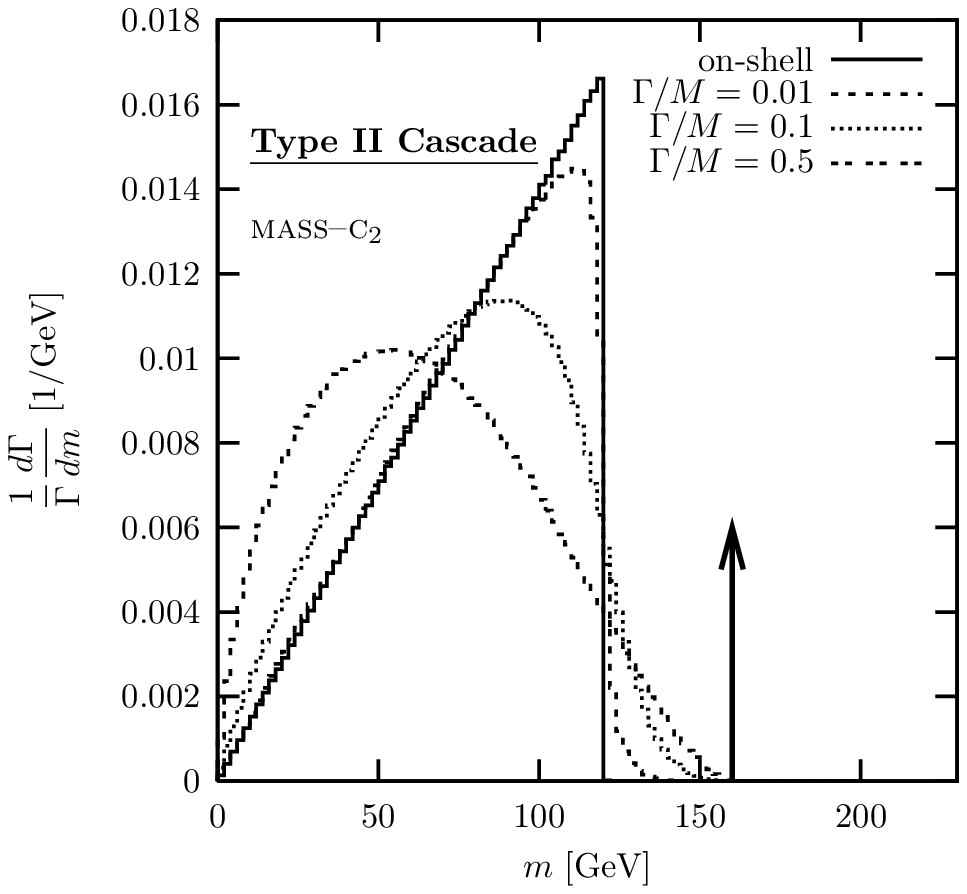}
  \caption{
Finite width effects on the normalized $m$ distribution.
We take the \mbo\;case for \tyo~decay, and \mct\;for \tyt~decay.
As before, we take
$\Gm/ M =0,~0.01,~0.1,~0.5$.
}
\label{fig:gm:m}
\end{figure}

We consider more general mass parameters,
\mbo\;for the \tyo~and \mct\;for the \tyt~cascade decay. 
First we examine the finite width effects on the invariant mass 
distributions in Fig.~\ref{fig:gm:m}.
These cases
show more stable cusp and endpoint structures from the finite width effects.
For $\Gm/M =1\%$, 
the $m$ distributions in both \tyo~and \tyt~decays
do not change, keeping the same cusp and endpoint structures.
For 10\% of $\Gm/M$, the $m$ cusp of the 
\tyo~decay retains its position,
though losing its sharpness.
The $m$ endpoints in both \tyo~and \tyt~decays
are shifted into the new position $m_D -2 m_X$.
If $\Gm/M = 50\%$,
the \tyo~decay does not retain the shape and position of the cusp,
and 
the \tyt~decay does not show the right-angled triangle shape of
the $m$ distribution.
Both cases have the same new endpoint
at $m_D - 2 m_X$,
which is also valuable information 
for the missing particle mass measurement.

\begin{figure}[!t]
\centering
  \includegraphics[scale=0.75]{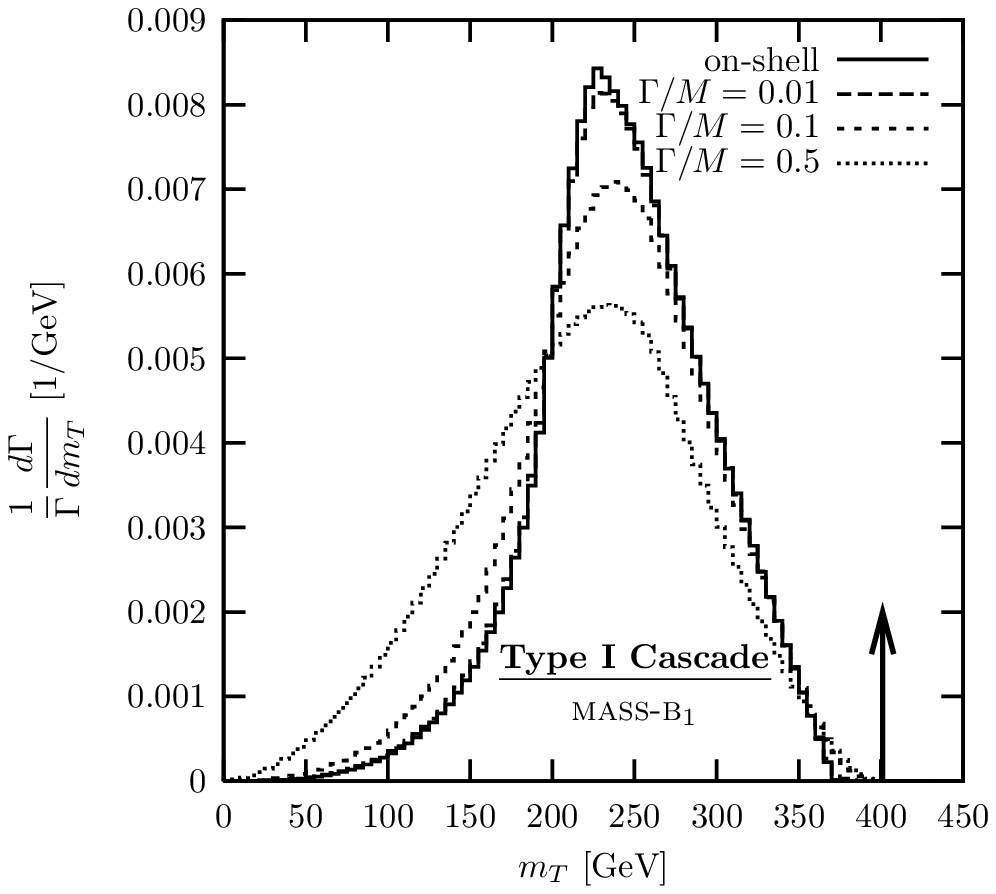}
\includegraphics[scale=0.75]{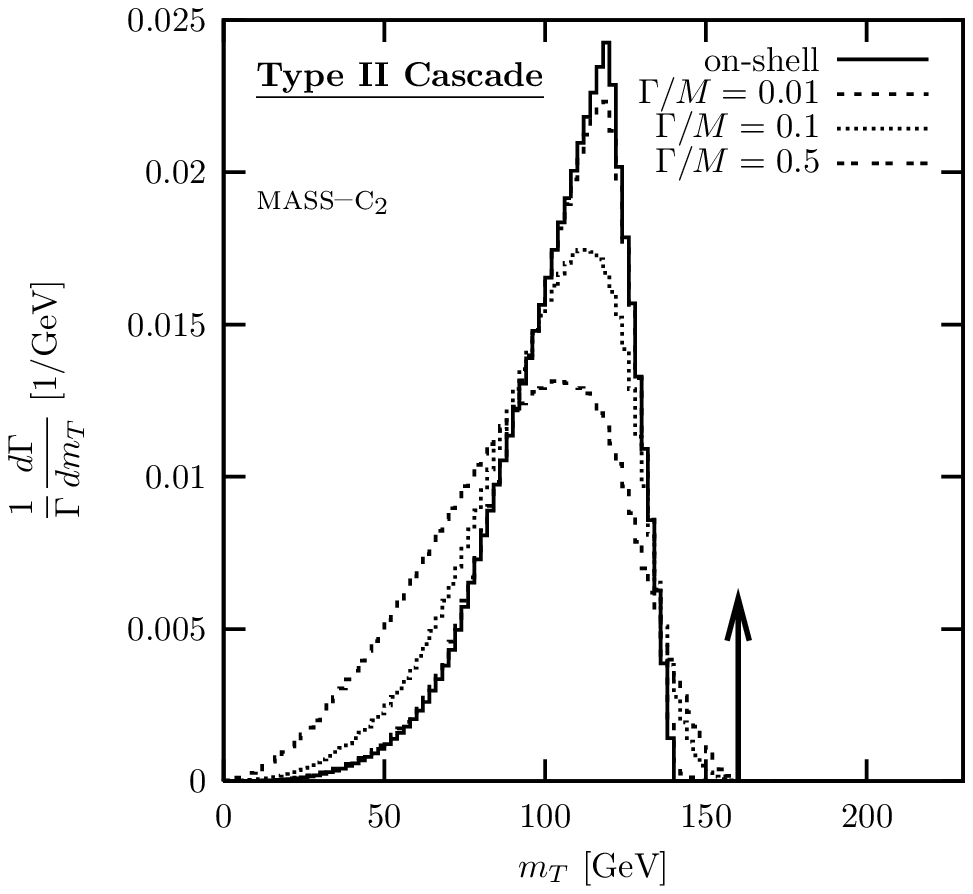}
  \caption{
The width effects on the normalized
$d \Gm /d m_T$
for the \tyo~\mbo\;and \tyt~\mct\;cascade decays.
As before, we take
$\Gm/ M =0,~0.01,~0.1,~0.5$.}
\label{fig:gm:mT}
\end{figure}

In Fig.~\ref{fig:gm:mT},
we show the width effects 
on the $m_T$ distributions.
The $m_T$ cusp structures are more stable than the $m$ cusps
in both \tyo~and
\tyt~decays.
For $\Gm/M =1\%$,
the changes in the distribution are unnoticeable.
For $\Gm/M \gsim 10\%$, 
we start to lose the sharpness of the cusps
but still keep the positions for the cusp in both cases.
If $\Gm/M = 50\%$, 
the cusped peaks become dull further with relatively stable positions,
and the $m_T$ endpoints are shifted into
$m_D - 2 m_X$.

\begin{figure}[!t]
\centering
  \includegraphics[scale=0.75]{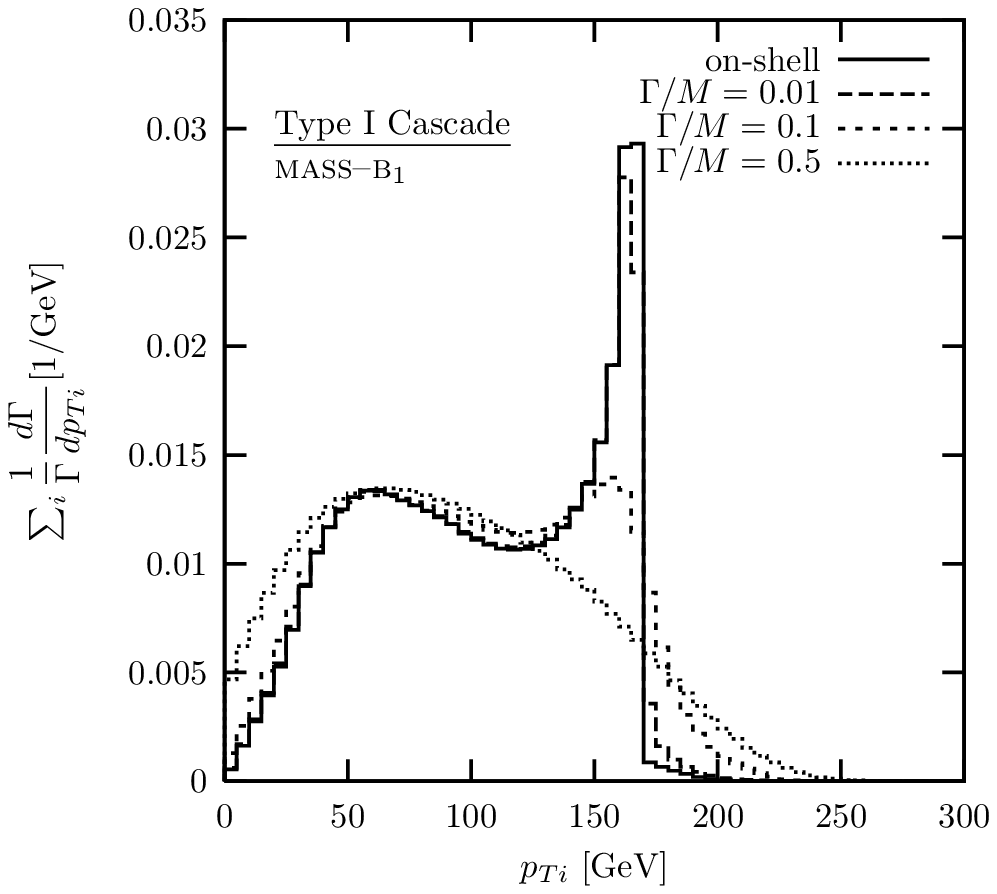}
\includegraphics[scale=0.75]{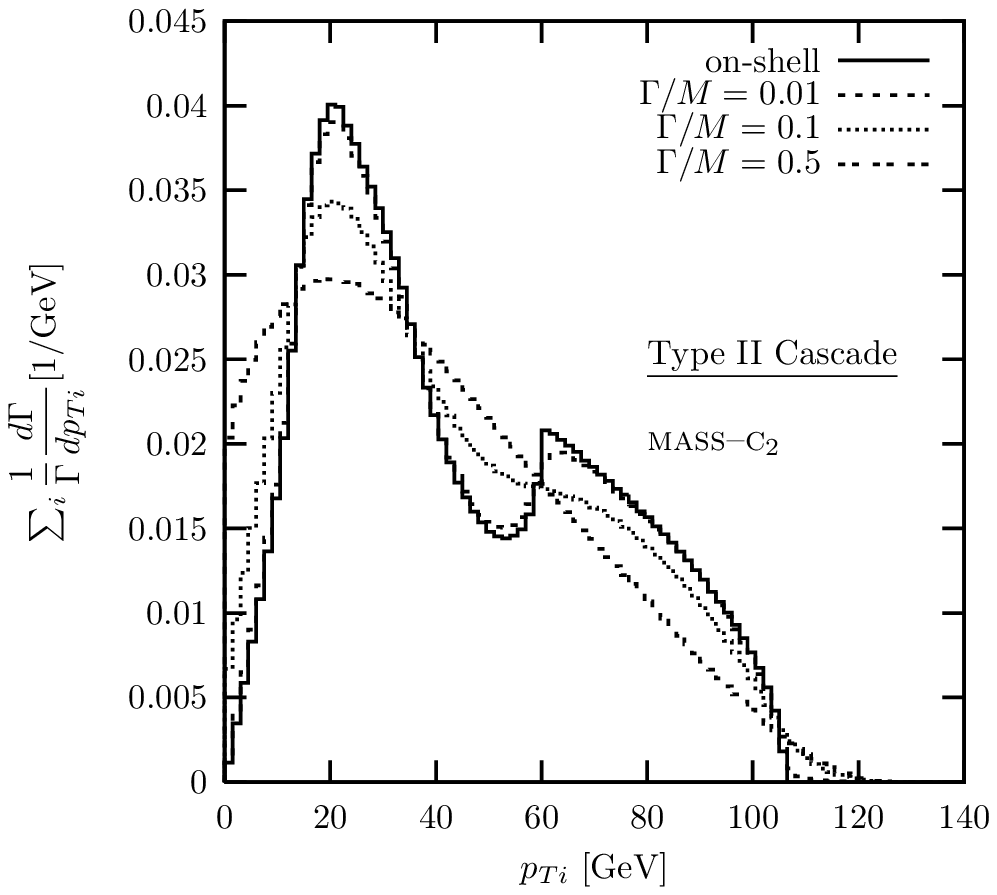}
  \caption{
The width effects on the summed  distributions of $p_{Ti}$
for $\Gm/ M =0,~0.01,~0.1,~0.5$.
}
\label{fig:gm:pTisum}
\end{figure}

\begin{figure}[!t]
\centering
  \includegraphics[scale=0.75]{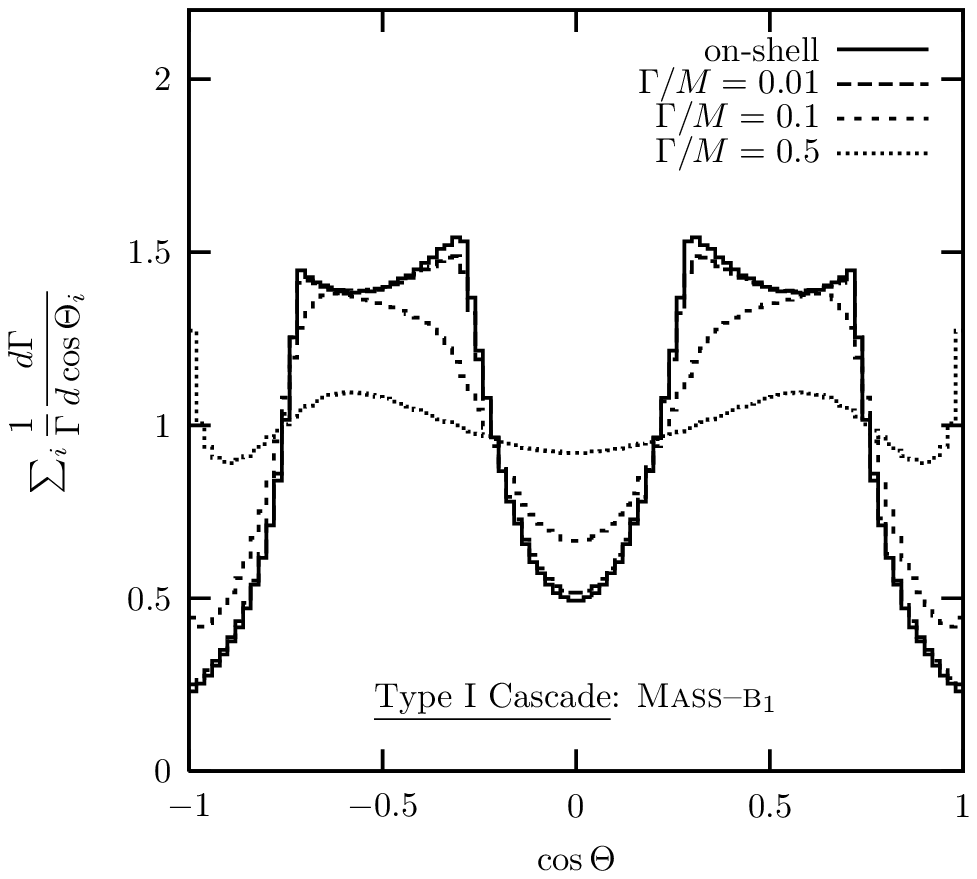}
\includegraphics[scale=0.75]{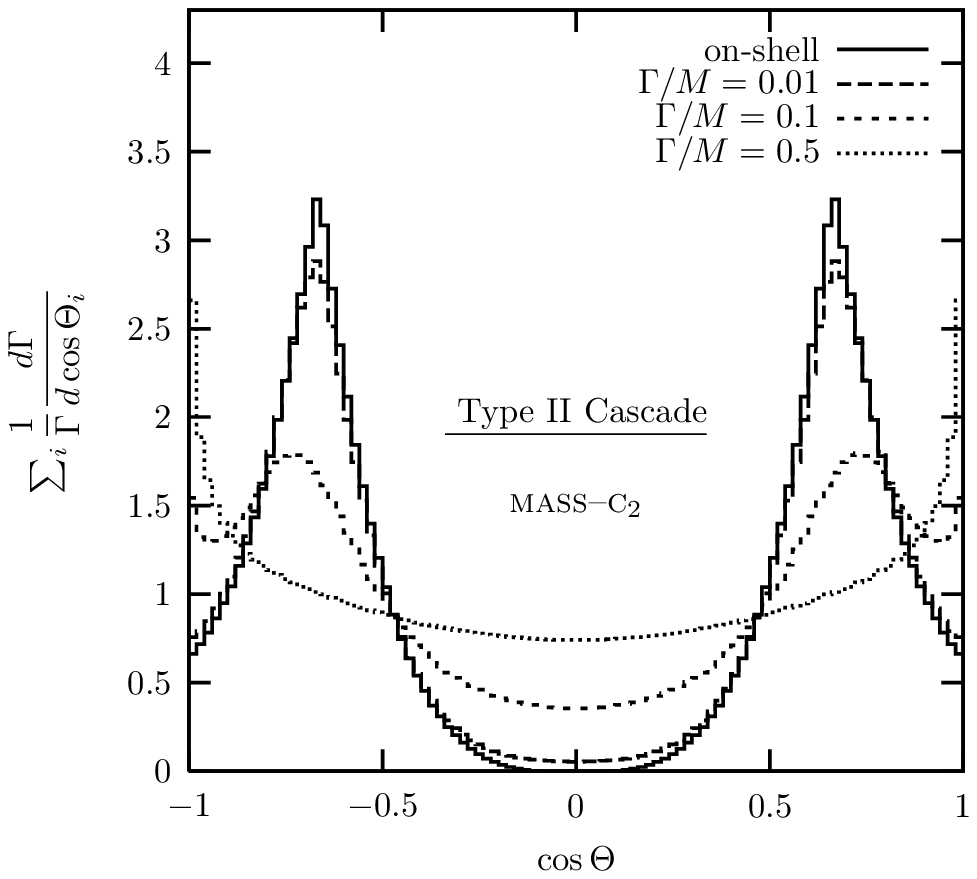}
  \caption{
The summed $\cos\Th$ distributions
for $\Gm/ M =0,~0.01,~0.1,~0.5$.
}
\label{fig:gm:costh}
\end{figure}

Figures \ref{fig:gm:pTisum} and \ref{fig:gm:costh}
show the width effects 
on the summed distributions of $p_{Ti}$ and $\cos\Th_i$ respectively.
Both distributions preserve the cusp structure 
for $\Gm/M =1\%$.
If $\Gm/M \gsim 10\%$, however,
the finite width effects almost smear
the cusp and endpoint structures.

\subsection{Longitudinal boost effect}

In hadronic collisions,
the longitudinal motion of the particle $D$ is not determined,
which affects only the 
$\cos\Th$.
The angle $\Th_i$ of the
visible particle $a_i$ is defined in the c.m. frame of $a_1$
and $a_2$ with respect to their c.m. moving direction,
and this direction is defined in the $D$ rest frame.

\begin{figure}[!t]
\centering
  \includegraphics[scale=0.75]{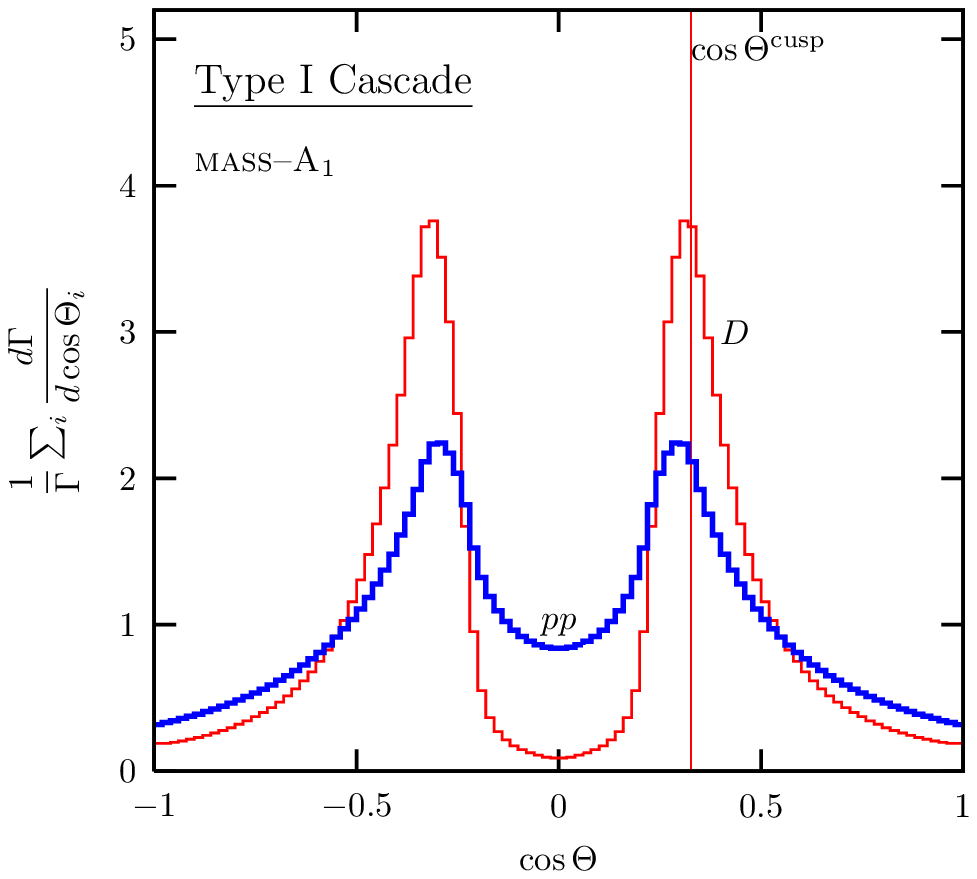}
\includegraphics[scale=0.75]{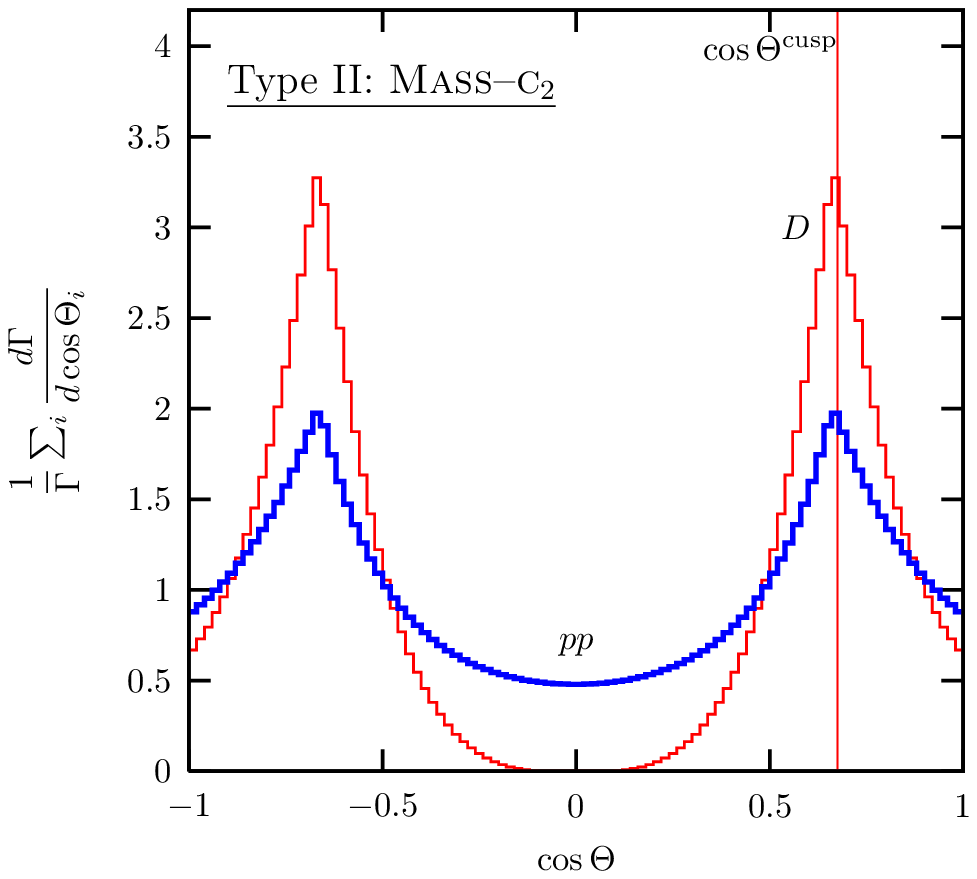}
  \caption[Normalized differential decay rates with respect to
$\cos\Theta$ ]
{
Normalized differential decay rates versus
$\cos\Theta$ in the $D$-rest frame (thin curves) and in
the $pp$ lab frame with $\sqrt{s}=14\ \tev$ (thick curves).}
\label{fig:longitudinal:boost}
\end{figure}

In order to see the longitudinal boost effects,
we convert the $\cos\Theta$ distribution in the $D$ rest frame into
the $pp$ frame at the LHC, 
by convoluting with the parton distribution
functions of a proton.
In Fig.~\ref{fig:longitudinal:boost},
we compare the summed distributions of $\cos\Th_i$
in the $D$-rest frame (thin curves) with that in
the $pp$ lab frame at $\sqrt{s}=14\ \tev$ (thick curves).
For the parton distribution function,
we have used
CTEQ6~\cite{cteq6}.
We take the \mao\;for \tyo~and the \mct\;for \tyt~decay.
For simplicity 
we assume that the heavy particle $D$ 
is singly produced through the $s$-channel gluon fusion 
and $q\bar{q}$ annihilation.

Unlike the finite width effects,
the longitudinal boost effect does not completely smash
the characteristic shape.
The sharp cusp structures survive to some extent
in both \tyo~and \tyt~cascade decays.
The shift of the $\cos\Th$ cusp position 
is minor.
Moreover the overall functional shape remains the same
even though the absence of events around $\cos\Th =0$ 
in the $D$ rest frame is filled by the longitudinal boost effects.
The cusp in the $\cos\Th$ distribution, though Lorentz non-invariant,
is quite useful to draw
mass information.
Again we emphasize that the $e^+ e^-$ linear collider does not have 
this unambiguity.

\subsection{Spin-correlation effect}
\label{subsec:spin}
\begin{figure}[!t]
\centering
\includegraphics[scale=0.9]{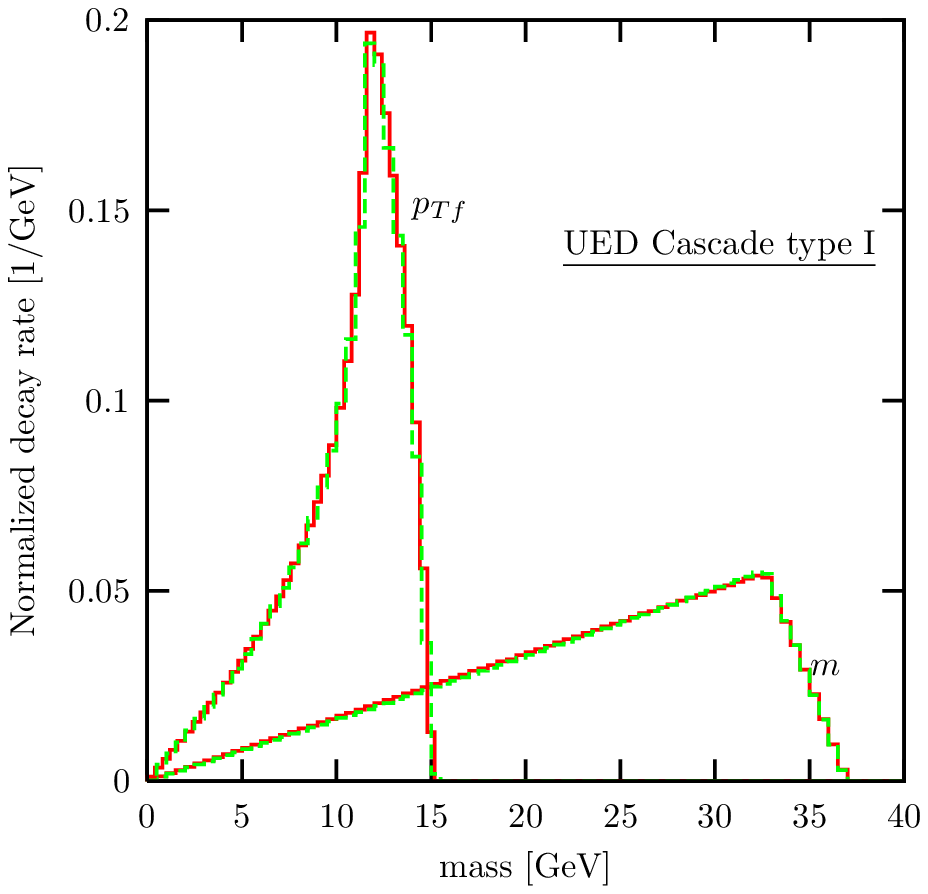}
  \caption{
The 
$d\Gm/d m$ and $d\Gm/d p_{Ti}$
for the process of $\zt \to \ell +\lt  \to \ell  + \lkp\lo
\to \ell \ell \lkp\lkp$ 
with and without spin correlations. }
\label{fig:spin:correlation}
\end{figure}

Our main results are based on the kinematics only,
ignoring the spin-correlation in the full matrix elements.
Since this paper is focused on the basic properties
of the kinematic singular structures in the cascade decays,
full analysis for each new physics process
is beyond the scope of this paper.
Nevertheless 
the algebraic singularity origin of the cusp and endpoint keeps
them stable under
the spin correlation effects~\cite{Kim:2009si}.

In order to demonstrate this,
we consider one example, the $\zt$ decay in the
the UED model:
\bea
\hbox{Cascade \tyo: ~~} && \zt \to \ell +\lt  \to \ell  + \lkp\lo
\to \ell \ell \lkp\lkp .
\eea
In Fig.~\ref{fig:spin:correlation},
we show their spin correlation effects.
We found that the spin correlations
do not change the $m$ and  $p_{Ti}$ distributions.
Two distributions with and without spin-correlation effects are almost
identical.

\section{Summary and Conclusions}
\label{sec:conclusions}
We have studied the singularity structure, 
such as cusps and endpoints, 
in the kinematic distributions
of three-step cascade decay 
of a new parity-even particle $D$
and the determination of 
 the missing particle mass 
by using such singularities.

Two non-trivial decay topologies,
called the \tyo~and \tyt~cascade decays, have been studied.
In the \tyo~decay ($D\to a_1 C,~C\to X_1 B, ~B\to a_2 X_2$), where
the first missing particle $X_1$ is from the second-step decay,
the distribution of the invariant mass $m$ 
of two visible particles, $a_1$ and $a_2$, develops a cusp.
Full functional form of the $m$ distribution for general
mass parameters has been derived.
If the mother particle $D$ is produced at rest in the transverse direction,
various longitudinal-boost invariant observables 
accommodate cusp structures.
First there is a cusp in the transverse mass $m_T$ distribution,
which is complementary for the $m$ cusp 
since the $m_T$ cusp shape is sharp even when the $m$ cusp is dull. 
Although the transverse momentum distribution of the c.m. system 
of two visible particles does not develop a visible cusp structure and 
a sharp endpoint, we note that 
the transverse momentum distribution of the far visible
particle $a_2$ has a cusp, and that of the near visible particle $a_1$
has an endpoint of the shape of a steep cliff.
We also study the summed
distribution of $\cos\Th_i$, which has two independent cusp structures.

\begin{table}[t!]
\centering
\begin{tabular}{|c||c|cc|}
\hline
 & \multirow{2}{*}{~~Antler~~} &  \multicolumn{2}{c|}{Cascade} \\ 
 &  & ~~\tyo~~ & ~~\tyt~~ \\ \hline
  $m$ & yes &  yes & no\\\hline
    $m_T$ & no & yes & yes \\\hline
  $p_T$ & no &  no & no \\\hline
  $p_{Tn}$ & \multirow{2}{*}{yes} &  no & yes \\
   $p_{Tf}$  &   & yes & no \\\hline
 ~~~$\cos\Theta$~~~ & yes & yes  & yes \\
\hline
\end{tabular}
\caption{ The presence or absence of the cusp
in the kinematic distributions of $m$, $m_T$, $p_T$, $p_{Ti}$, $\cos\Theta$
of the antler, \tyo~cascade, \tyt~cascade decays.}
\label{table:summary}
\end{table}

In the \tyt~decay ($D \to X_1 C, ~C\to a_1 B, ~B\to a_2 X_2$),
the first missing particle $X_1$ is from the first step decay.
The kinematics of 
the two visible $a_1$ and $a_2$
is determined solely by the two-step cascade decay from the first intermediate 
particle $C$,
so that the invariant mass distribution 
does not have a cusp structure.
However,  the kinematic distributions
regarding the transverse motion
 from production of both $X_1$ and $X_2$ 
can carry the information from the whole three-step cascade decay.
We show both the $m_T$ and $\sum_i p_{Ti}$ 
distributions have distinctive cusp structures.
In the individual transverse momentum distribution,
only the near visible particle has both a sharp cusp
and a fast-falling endpoint.
The $\cos\Th$ distribution also shows  
a cusp as well.
Including the antler decay topology,
we have summarized the existence of cusp 
in the kinematic distributions of $m$, $m_T$, $p_T$, $p_{Ti}$, $\cos\Theta$
in Table \ref{table:summary}.

We have also considered the effects of 
finite decay widths, longitudinal-boost of the parent particle $D$,
and spin correlation. 
The effects of the finite widths of the intermediate particles 
can be significant 
if the decay width is 
sizable, say $\Gm/M \gsim 10\%$.
As the decay width increases,
the sharp cusp gets smeared, and the endpoint position
gets shifted to $m_D -2 m_X$:
the missing particle mass $m_X$ can be 
still extracted by a proper fitting.
The longitudinal motion of the parent 
particle $D$ affects the distribution of $\cos\Th$.
At least for the sample mass parameters, however,
the $\cos\Th$ cusp remains sharp after convoluting with the parton distribution functions
of a proton at the LHC.
Spin correlation effects 
from full $S$-matrix elements 
turn out to be negligible in most cases,
which is expected since the singularities are determined 
by the kinematic relations.

With the companion paper on the detailed study of the kinematic cusps in the 
antler decay~\cite{long:antler}, our analysis shows the general kinematic properties and 
provides useful formulae for the decay topologies with 
two visible particles and two missing particles. By looking at the singularity
structures of various kinematic distributions, the hidden nature of the 
missing particle can be probed effectively and elegantly. 
With the outstanding performance of the LHC and detectors, 
this is an exciting time for such investigation.

\acknowledgments
This work was supported in part by the U.S.~Department of Energy under grant No. DE-FG02-12ER41832., and in part by PITT PACC. The work of JS was supported by WCU program
through the NRF funded by the MEST (R31-2008-000-10057-0).

\appendix*

\section{Invariant mass distributions for the general \tyo~case}
\label{appendix:inv:mass}

In this appendix,
we present the invariant mass distribution in
the general \tyo~cascade decays:
\bea
D(P) \longrightarrow  && C + a_1 (k_1),       \\
\no
          && C \longrightarrow       B + X_1 ,  \\ \no
          &&  \hspace{34pt}
 B \longrightarrow      a_2(k_2) + X_2.
\eea
As discussed in the main text, the \tyt~cascade decay is practically 
a three body decay in the view point of visible particles.
This four-body decay has generally seven different mass parameters.
We define the rapidities of six particles as
\begin{eqnarray}
\cosh \xi_{C} &=& 
\frac{m_D^2+ m_C^2 - m_{a_1}^2}{2 m_D m_C}, \qquad 
\cosh \xi_{a_1} =
\frac{m_D^2 + m_{a_1}^2 - m_{C}^2}{2 m_D m_{a_1}}, 
\\ \no
\cosh \xi_{B} &=&
\frac{m_C^2 + m_B^2 - m_{X_1}^2 }{2 m_C m_B}, \qquad
\cosh \xi_{X_1} = 
\frac{m_C^2 + m_{X_1}^2 - m_{B}^2 }{2 m_C m_{X_1}},
 \\ \no
\cosh \xi_{a_2} &=&
\frac{m_B^2 + m_{a_2}^2 - m_{X_2}^2}{2 m_B m_{a_2}}, \qquad
\cosh \xi_{X_2} =
\frac{m_B^2 + m_{X_2}^2 - m_{a_2}^2 }{2 m_B m_{X_2}}.
\end{eqnarray}
A very useful kinematic  variable is $\chi$,
the rapidity of the particle $a_2$ in the rest frame of $a_1$:
\begin{eqnarray}
\chi \equiv \cosh \xi^{(a_1)}_{a_2} = 
\frac{m^2-m_{a_1}^2 - m_{a_2}^2}{2 m_{a_1} m_{a_2}},
\end{eqnarray}
where the superscript $(a_1)$
denotes that the rapidity is defined in 
the rest frame of $a_1$.

The functional expression
of $d \Gamma / d m$ is different according to the 
mass relations.
The derivation of $d \Gamma / d m$ is similar to that
presented in the appendix of Ref.~\cite{long:antler}.
For simple presentation,
we introduce
\begin{eqnarray}
\xi_{++} &=& \xi_{B} + \xi_{a_1} + \xi_{a_2} + \xi_{C}, \\
\xi_{+-} &=& |\xi_{B}+\xi_{a_1}-\xi_{a_2}-\xi_{C}|,  \\
\xi_{-+} &=& |\xi_{B}-\xi_{a_1}+\xi_{a_2}+\xi_{C}|, \\
\xi_{--} &=& |\xi_{B}-\xi_{a_1}-\xi_{a_2}-\xi_{C}|.
\end{eqnarray}
We order $\xi_{+-}$, $\xi_{-+}$ and $\xi_{--}$ 
and name them 
$\xi_1 \leq \xi_2 \leq \xi_3$. 
Analytic functions forms of ${d\Gamma}/{d\chi}$
are then written as
\begin{itemize}
\item if
$|\xi_{B} - \xi_{a_2} - \xi_{C}| \geq \xi_{a_1}$ or 
$\xi_{B} + \xi_{a_2} + \xi_{C} \leq \xi_{a_1}$, 
\begin{eqnarray}
\label{eq:3piece}
\frac{d\Gamma}{d\chi} \propto
\begin{cases} 
\begin{array}{ll}
-\xi_1 + \cosh^{-1} \chi, &
\mbox{ if $\cosh \xi_1 \leq \chi \leq \cosh \xi_2$, } \\
\xi_2 -\xi_1,  &
\mbox{ if $\cosh \xi_2 \leq \chi \leq \cosh \xi_3$, } \\
\xi_{++} - \cosh^{-1} \chi,  &
\mbox{ if $\cosh \xi_3 \leq \chi \leq \cosh \xi_{++}$, } \\
0, & \mbox{ otherwise.}
\end{array}
\end{cases}
\end{eqnarray}
\item if
$|\xi_{B}-\xi_{a_2}-\xi_{C}| < \xi_{a_1} 
< \xi_{B}+\xi_{a_2}+\xi_{C}$, 
\begin{eqnarray}
\label{eq:4piece}
\frac{d\Gamma}{d\chi} \propto
\begin{cases} 
\begin{array}{ll}
2 \cosh^{-1} \chi, & \mbox{ if $1 \leq \chi \leq \cosh \xi_1$, } \\
\xi_1 + \cosh^{-1} \chi, &
\mbox{ if $\cosh \xi_1 \leq \chi \leq \cosh \xi_2$, } \\
\xi_1 + \xi_2,  &
\mbox{ if $\cosh \xi_2 \leq \chi \leq \cosh \xi_3$, } \\
\xi_{++} - \cosh^{-1} \chi,  &
\mbox{ if $\cosh \xi_3 \leq \chi \leq \cosh \xi_{++}$, } \\
0, & \mbox{ otherwise.}
\end{array}
\end{cases}
\end{eqnarray}
\end{itemize}

The positions of the minimum,
cusp, and maximum of the invariant mass distribution
are
\bea
\label{eq:min:cusp:max:geneal}
M^{\rm min}_{\rm cas1} &=&
\left\{
\renewcommand{\arraystretch}{1.3}
\begin{array}{ll}
\sqrt{m_{a_1}^2+m_{a_2}^2 + 2 m_{a_1} m_{a_2}\cosh \xi_1},
& \hbox{ for } \mathcal{R}_{1,\cdots,6} \\
m_{a_1}+m_{a_2},
& \hbox{ for } \mathcal{R}_{7,\cdots,12} \\
\end{array}
\right.
\\ \no
M^{\rm cusp}_{\rm cas1} &=& 
\sqrt{m_{a_1}^2+m_{a_2}^2 + 2 m_{a_1} m_{a_2}\cosh \xi_3},
\\ \no
M^{\rm max}_{\rm cas1} &=& 
\sqrt{m_{a_1}^2+m_{a_2}^2 + 2 m_{a_1} m_{a_2}\cosh \xi_{++}}.
\eea

\def\PRD #1 #2 #3 {Phys. Rev. D {\bf#1},\ #2 (#3)}
\def\PRL #1 #2 #3 {Phys. Rev. Lett. {\bf#1},\ #2 (#3)}
\def\PLB #1 #2 #3 {Phys. Lett. B {\bf#1},\ #2 (#3)}
\def\NPB #1 #2 #3 {Nucl. Phys. B {\bf #1},\ #2 (#3)}
\def\ZPC #1 #2 #3 {Z. Phys. C {\bf#1},\ #2 (#3)}
\def\EPJ #1 #2 #3 {Euro. Phys. J. C {\bf#1},\ #2 (#3)}
\def\JHEP #1 #2 #3 {JHEP {\bf#1},\ #2 (#3)}
\def\IJMP #1 #2 #3 {Int. J. Mod. Phys. A {\bf#1},\ #2 (#3)}
\def\MPL #1 #2 #3 {Mod. Phys. Lett. A {\bf#1},\ #2 (#3)}
\def\PTP #1 #2 #3 {Prog. Theor. Phys. {\bf#1},\ #2 (#3)}
\def\PR #1 #2 #3 {Phys. Rep. {\bf#1},\ #2 (#3)}
\def\RMP #1 #2 #3 {Rev. Mod. Phys. {\bf#1},\ #2 (#3)}
\def\PRold #1 #2 #3 {Phys. Rev. {\bf#1},\ #2 (#3)}
\def\IBID #1 #2 #3 {{\it ibid.} {\bf#1},\ #2 (#3)}

\end{document}